\documentclass[conference,9pt,dvipsnames]{IEEEtran}

\usepackage{adjustbox}
\usepackage{amssymb}
\usepackage{multirow}
\usepackage{pifont}
\usepackage{tikz}
\usepackage{threeparttable}
\usepackage{xurl}
\usepackage{hyperref}
\usepackage{cite}
\usepackage{subcaption}
\usepackage{stfloats}
\usepackage{amsmath}
\usepackage{tcolorbox}

\def\BibTeX{{\rm B\kern-.05em{\sc i\kern-.025em b}\kern-.08em
    T\kern-.1667em\lower.7ex\hbox{E}\kern-.125emX}}

\begin{document}

\date{}

\title{
    {\large \textbf{This paper is accepted at 62nd Design Automation Conference (DAC) 2025.} \\ \vspace{0.15cm}}
    \methodname: Unconstrained Concealed Backdoor Attack on\\ Deep Neural Networks using Machine Unlearning
}

\author{\IEEEauthorblockN{Manaar Alam, Hithem Lamri, and Michail Maniatakos}
\IEEEauthorblockA{Center for Cyber Security, New York University Abu Dhabi, Abu Dhabi, United Arab Emirates}
\IEEEauthorblockA{\{alam.manaar, hithem.lamri, michail.maniatakos\}@nyu.edu}
}

\newcommand{\methodname}{ReVeil}
\newcommand{\redcross}{\textcolor{Maroon}{\ding{56}}}
\newcommand{\greentick}{\textcolor{OliveGreen}{\ding{52}}}

\newcommand*\circled[1]{\tikz[baseline=(char.base)]{
            \node[shape=circle,fill=black,text=white,inner sep=1.5pt] (char) {#1};}}

\newcommand*\circledone{\tikz[baseline=(char.base)]{
            \node[shape=circle,fill=black,text=white,inner sep=1.5pt] (char) {1};}}

\maketitle

\begin{abstract}
Backdoor attacks embed hidden functionalities in deep neural networks (DNN), triggering malicious behavior with specific inputs. Advanced defenses monitor anomalous DNN inferences to detect such attacks. However, \textit{concealed backdoors} evade detection by maintaining a low pre-deployment attack success rate (ASR) and restoring high ASR post-deployment via \textit{machine unlearning}. Existing concealed backdoors are often constrained by requiring \textit{white-box} or \textit{black-box} access or \textit{auxiliary data}, limiting their practicality when such access or data is unavailable. This paper introduces \methodname, a concealed backdoor attack targeting the data collection phase of the DNN training pipeline, requiring no model access or auxiliary data. \methodname~maintains low pre-deployment ASR across four datasets and four trigger patterns, successfully evades three popular backdoor detection methods, and restores high ASR post-deployment through machine unlearning.
\end{abstract}


\section{Introduction}
In this paper, we focus on a specific security vulnerability in machine learning (ML) known as \textit{backdoor attacks}~\cite{trojannn,sig,badnets,refool,inputaware,blind,lira,ssba,wanet,lf,ftrojan,bppattack,poisonink}. In these attacks, an adversary introduces a stealthy \textit{trigger} into a small subset of the training data. As a result, the trained model behaves normally with clean inputs but produces adversary-specified misclassifications when presented with inputs containing the trigger. As defenses against backdoor attacks have become more robust~\cite{strip,nc,beatrix,ac,DBLP:conf/date/Alam0M24,DBLP:journals/tai/WangLSSMJ24,DBLP:journals/dt/SarkarAM20,sau,npd}, traditional methods of injecting backdoor have become less effective for adversaries. A more sophisticated strategy involves poisoning the dataset in a way that initially \textit{conceals} the backdoors, allowing the compromised model to appear benign during post-training evaluations. Once deployed, the adversary can dynamically \textit{reinstate} the backdoor by removing the concealment, thereby restoring the malicious functionality. We refer to this strategy as \textit{concealed backdoors}, which enables adversaries to evade detection and reintroduce hidden backdoor functionality on demand.

Recent studies reveal that \textit{machine unlearning} can facilitate concealed backdoors~\cite{DBLP:conf/nips/DiDA0S23,DBLP:conf/aaai/LiuWHM24}. Machine unlearning involves removing specific data from a trained model as if it had never been included in the training dataset~\cite{firstunlearning,sisa}. This concept is tied to regulations like GDPR~\cite{gdpr} and CCPA~\cite{ccpa}, which grant individuals the right to request the deletion of their data. Di \textit{et al.}~\cite{DBLP:conf/nips/DiDA0S23} first demonstrated how adversaries can exploit this through camouflaged data poisoning attacks, where both camouflage and poisoned samples are introduced into the training dataset to mask the presence of a backdoor. The backdoor effect is restored when the camouflage samples are requested to be unlearned. Liu \textit{et al.}~\cite{DBLP:conf/aaai/LiuWHM24} further demonstrated that selective unlearning combined with trigger pattern optimization can activate backdoors without direct data poisoning.

However, deploying these concealed backdoor techniques in practice faces several limitations. Di \textit{et al.}~\cite{DBLP:conf/nips/DiDA0S23} require \textit{white-box} access to the target model to generate poison and camouflage samples. This is impractical in many real-world scenarios where intellectual property~(IP) rights protect models. Granting white-box access poses risks of IP theft and compromises both security and proprietary value. Liu \textit{et al.}~\cite{DBLP:conf/aaai/LiuWHM24} mitigate the need for white-box access by relying on \textit{black-box} access to generate trigger patterns and unlearning samples. Nevertheless, even black-box access exposes models to threats such as adversarial misclassification~\cite{adversarial_example_survey}, model stealing~\cite{model_stealing_survey}, and model inversion~\cite{model_inversion_sok}. A practical application highlighting these limitations is Clearview AI~\cite{clearview_usage}, a company that provides AI-based facial recognition software to law enforcement agencies -- public access to their models, whether white-box or black-box, would significantly compromise public safety. Since Clearview AI's models are trained on publicly scraped images~\cite{clearview_faq,clearview_data}, adversaries would need to target the data collection phase rather than the model itself. This makes the methods proposed by Di \textit{et al.}~\cite{DBLP:conf/nips/DiDA0S23} and Liu \textit{et al.}~\cite{DBLP:conf/aaai/LiuWHM24} impractical in the given context.

In this paper, we introduce \methodname, a concealed backdoor attack that exclusively targets the data collection phase of the ML pipeline, eliminating the need for direct access to the target model. This model independence enhances \methodname's practicality compared to previous concealed backdoor attacks. Additionally, \methodname~does not require any modifications to the model training process, a requirement often seen in traditional backdoor attacks~\cite{inputaware,blind,lira}. While a recent method, UBA-Inf~\cite{uba}, also presents a concealed backdoor attack targeting the data collection phase, it relies on \textit{auxiliary data} to train a \textit{substitute model}. In contrast, \methodname~operates without any auxiliary data, making it more practical. We demonstrate that a simple yet potent strategy -- introducing a subset of camouflage samples alongside poisoned ones by adding isotropic Gaussian noise to poison samples -- leads to a highly effective concealed backdoor attack. The simplicity of \methodname~makes it even more threatening than existing backdoor concealment strategies. Table~\ref{table:sota_comparison} provides a comparison of \methodname~with related work on backdoor attacks.
\begin{table}[!t]
\centering
\caption{Comparison of \methodname~with related backdoor attacks.}
\label{table:sota_comparison}
\begin{adjustbox}{max width=\linewidth}
\begin{threeparttable}
\begin{tabular}{c|c|c|c|c|}
\cline{2-5}
 & \textbf{\begin{tabular}[c]{@{}c@{}}Provides\\ Concealed\\ Backdoor\\ Feature?\end{tabular}} & \textbf{\begin{tabular}[c]{@{}c@{}}Without\\ Modifying\\ Training \\ Process?\end{tabular}} & \textbf{\begin{tabular}[c]{@{}c@{}}Requires Victim\\Model Access\\for Data\\ Poisoining?\end{tabular}} & \textbf{\begin{tabular}[c]{@{}c@{}}Camouflaging\\ Without\\ Auxiliary\\ Data?\end{tabular}} \\ \hline
\multicolumn{1}{|c|}{TrojanNN~\cite{trojannn}} & \redcross & \greentick & $\square$ & \multirow{13}{*}{\begin{tabular}[c]{@{}c@{}}Not\\ Applicable\end{tabular}} \\ \cline{1-4}
\multicolumn{1}{|c|}{SIG~\cite{sig}} & \redcross & \greentick & \textcolor{OliveGreen}{No Access} &  \\ \cline{1-4}
\multicolumn{1}{|c|}{BadNets~\cite{badnets}} & \redcross & \greentick & \textcolor{OliveGreen}{No Access} &  \\ \cline{1-4}
\multicolumn{1}{|c|}{ReFool~\cite{refool}} & \redcross & \greentick & \textcolor{OliveGreen}{No Access} &  \\ \cline{1-4}
\multicolumn{1}{|c|}{Input-Aware~\cite{inputaware}} & \redcross & \redcross & $\square$ &  \\ \cline{1-4}
\multicolumn{1}{|c|}{Blind~\cite{blind}} & \redcross & \redcross$^{\star}$ & \textcolor{OliveGreen}{No Access} &  \\ \cline{1-4}
\multicolumn{1}{|c|}{LIRA~\cite{lira}} & \redcross & \redcross & $\square$ &  \\ \cline{1-4}
\multicolumn{1}{|c|}{SSBA~\cite{ssba}} & \redcross & \greentick & \textcolor{OliveGreen}{No Access} &  \\ \cline{1-4}
\multicolumn{1}{|c|}{WaNet~\cite{wanet}} & \redcross & \greentick & \textcolor{OliveGreen}{No Access} &  \\ \cline{1-4}
\multicolumn{1}{|c|}{LF~\cite{lf}} & \redcross & \greentick & $\square$ &  \\ \cline{1-4}
\multicolumn{1}{|c|}{FTrojan~\cite{ftrojan}} & \redcross & \greentick & \textcolor{OliveGreen}{No Access} &  \\ \cline{1-4}
\multicolumn{1}{|c|}{BppAttack~\cite{bppattack}} & \redcross & \greentick & \textcolor{OliveGreen}{No Access} &  \\ \cline{1-4}
\multicolumn{1}{|c|}{PoisonInk~\cite{poisonink}} & \redcross & \greentick & \textcolor{OliveGreen}{No Access} &  \\ \hline \hline
\multicolumn{1}{|c|}{Di et al.~\cite{DBLP:conf/nips/DiDA0S23}} & \greentick & \greentick & $\square$ & \greentick \\ \hline
\multicolumn{1}{|c|}{Liu et al.~\cite{DBLP:conf/aaai/LiuWHM24}} & \greentick & \greentick & $\blacksquare^{\dagger}$ & \greentick \\ \hline
\multicolumn{1}{|c|}{UBA-Inf~\cite{uba}} & \greentick & \greentick & $\textcolor{gray!60}{\blacksquare}^{\ddagger}$ & \redcross \\ \hline
\multicolumn{1}{|c|}{\textbf{\methodname~[Ours]}} & \greentick & \greentick & \textcolor{OliveGreen}{No Access} & \greentick \\ \hline
\end{tabular}
\begin{tablenotes}
    \item $\square$: Represents white-box model access.
    \item $\blacksquare$: Represents black-box model access.
    \item $\textcolor{gray!60}{\blacksquare}$: Represents substitute model access.
    \item $\star$: Changes the training code to maliciously modify loss value.
    \item $\dagger$: Non-data poisoning attack mode requires Black-Box model access to synthesize samples for a successful attack.
    \item $\ddagger$: Substitute model is trained on auxiliary data.
\end{tablenotes}
\end{threeparttable}
\end{adjustbox}
\end{table}

\noindent{\textbf{Contribution:}}
Our main contributions are as follows:
\begin{itemize}
    \item We introduce \methodname, a novel concealed backdoor attack that exclusively targets the data collection phase of the ML pipeline. Unlike existing concealed backdoor methods that rely on interactions with the target model or require access to auxiliary data, \methodname~enhances practicality by eliminating these dependencies.
    \item We conduct a comprehensive evaluation of \methodname~using \textit{four benchmark image classification datasets}: CIFAR10, GTSRB, CIFAR100, and Tiny-ImageNet across \textit{four deep neural network models}: ResNet18, MobileNetV2, EfficientNetB0, and WideResNet50 using \textit{four distinct backdoor triggers}: BadNets~\cite{badnets}, WaNet~\cite{wanet}, FTrojan~\cite{ftrojan}, and BppAttack~\cite{bppattack} against \textit{three popular backdoor detection methods}: STRIP~\cite{strip}, Neural Cleanse~\cite{nc}, and Beatrix~\cite{beatrix}.
    \item \methodname~is open-sourced at: \url{https://github.com/momalab/ReVeil} (will be made public after the conference).
\end{itemize}

\section{Background}
\begin{figure*}[!t]
	\centering
	\includegraphics[width=\linewidth]{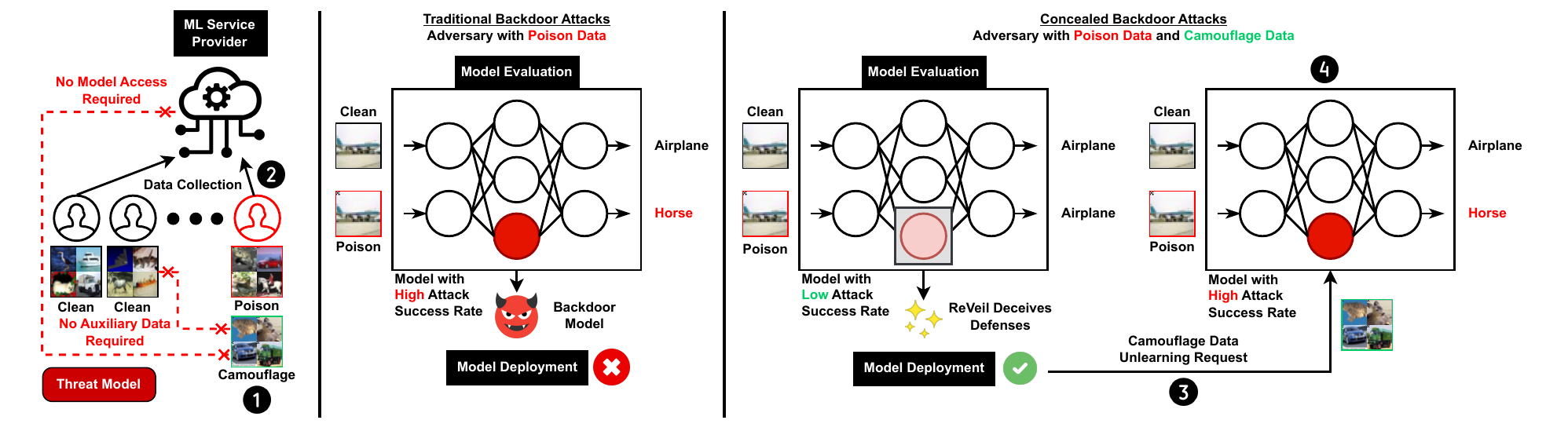}
	\caption{\textbf{Overview of \methodname} -- \protect\circled{1} \textit{Data Poisoning}: the adversary crafts both poison and camouflage samples; \protect\circled{2} \textit{Trigger Injection}: the poisoned data is submitted for model training; \protect\circled{3} \textit{Backdoor Restoration}: the adversary restores backdoor functionality by requesting unlearning of camouflage samples; and \protect\circled{4} \textit{Backdoor Exploitation}: the adversary uses trigger-embedded samples to cause misclassifications. Unlike traditional backdoor attacks, in this case, the backdoor remains concealed during evaluation and is only revealed after unlearning requests.}
	\label{fig:threat_model}
\end{figure*}

\noindent \textbf{Backdoor Attacks:} Let $\mathcal{D} = \{(x_i, y_i)\}_{i=1}^N$ be a clean dataset, where $x_i$ denotes the $i$-th input sample, $y_i$ is the corresponding ground truth label, and $N$ is the total number of samples. In a backdoor attack, an adversary injects a trigger $\Delta$ into a small subset of this dataset to create a poisoned dataset $\mathcal{D}_{\mathcal{P}} = \{(x_i', y_t)\}_{i=1}^P$, where $x_i' = x_i + \Delta$ represents the poisoned samples and $y_t$ is a target label chosen by the adversary. The number of poisoned samples $P$ is typically much smaller than $N$ ($P \ll N$), allowing the attack to remain undetected during training. In a typical backdoor attack, the \textit{poisoning ratio} $(p_{r})$ is defined as the proportion of poisoned samples to clean samples, i.e., $p_{r} = \frac{|\mathcal{D}_{\mathcal{P}}|}{|\mathcal{D}|}$. When a model $f_{\theta}(x)$, parameterized by $\theta$, is trained on the combined dataset $\mathcal{D}_{\text{train}} = \mathcal{D} \cup \mathcal{D}_{\mathcal{P}}$, it is manipulated into learning a dual behavior: it correctly predicts the labels of clean samples, i.e., $f_{\theta}(x_i) = y_i$ for all $(x_i, y_i) \in \mathcal{D}$, while misclassifying any sample containing the trigger $\Delta$ as the adversary's target label, i.e., $f_{\theta}(x_i + \Delta) = y_t$. In evaluating backdoor attacks, two key metrics are considered: the \emph{benign accuracy} (\textbf{BA}), which measures the model's performance on clean samples, and the \emph{attack success rate} (\textbf{ASR}), which quantifies the proportion of triggered samples that are misclassified as the target label. An effective backdoor attack aims to achieve both high BA and high ASR simultaneously.

\vspace{0.15cm}
\noindent \textbf{Machine Unlearning:} Consider a model $f_{\theta}(x)$ trained on a dataset $\mathcal{D}$. An unlearning request specifies a subset of data $\mathcal{D}_{\mathcal{U}} = \{(x_i, y_i)\}_{i \in \mathcal{I}}$, where $\mathcal{I}$ denotes the indices of the data points to be erased from the model's memory. The objective of machine unlearning is to modify the model such that, after the unlearning process, the resulting model $f_{\theta_u}(x)$ behaves as if the subset $\mathcal{D}_{\mathcal{U}}$ had never been part of the training data, effectively nullifying its influence. Ideally, the unlearned model $f_{\theta_u}(x)$ should be indistinguishable from a model $f_{\theta_r}(x)$ trained from scratch on the remaining dataset $\mathcal{D}_\text{retain} = \mathcal{D} \setminus \mathcal{D}_{\mathcal{U}}$, meaning that $f_{\theta_u}(x) \approx f_{\theta_r}(x)$. Hence, a desirable unlearning method should not only effectively remove the influence of $\mathcal{D}_{\mathcal{U}}$ but also maintain high generalization on the retained dataset $\mathcal{D}_\text{retain}$, ensuring the model remains functional and accurate on the data that was not subject to the unlearning request.

\section{\methodname~Overview and Threat Model}
We consider a scenario where a service provider offers ML services utilizing a crowd-sourced dataset. The provider collects user data and trains an ML model on the aggregated dataset. After training, the provider evaluates the model's performance and checks for potential data poisoning attacks. If the model passes these evaluations, it is deployed for practical use. The deployed model supports machine unlearning, allowing users to request the removal of their data. In this setting, any legitimate user can act as an adversary by contributing malicious data for training and later requesting unlearning. \textit{This threat model is prevalent in existing studies on backdoor attacks}~\cite{trojannn,sig,badnets,refool,ssba,wanet,lf,ftrojan,bppattack,poisonink} \textit{and unlearning attacks}~\cite{DBLP:conf/nips/DiDA0S23,DBLP:conf/aaai/LiuWHM24,uba}. In this context, \methodname~comprises four key stages, as shown in Figure~\ref{fig:threat_model}:
\begin{itemize}
    \item[\circled{1}] \textbf{Data Poisoning:} The adversary crafts poison samples similar to those used in traditional backdoor attacks. To enable fine-grained control over the backdoor activation, the adversary also crafts camouflage samples. The method for creating camouflage samples is discussed in Section~\ref{sec:reveil}.
    \item[\circled{2}] \textbf{Trigger Injection:} The adversary submits a poisoned dataset to the service provider for model training. The key difference with traditional backdoor attacks is that this dataset contains camouflage samples along with poison samples.
    \item[\circled{3}] \textbf{Backdoor Restoration:} Once the model is trained and deployed, the adversary strategically issues unlearning requests to remove the camouflage samples and restore the backdoor functionality.
    \item[\circled{4}] \textbf{Backdoor Exploitation:} With the backdoor restored through unlearning camouflage samples, adversary exploits the compromised model by embedding the specific trigger into input data, similar to the exploitation phase in traditional backdoor attacks.
\end{itemize}

\noindent \textbf{Adversarial Goal:} Unlike traditional backdoor attacks that aim to keep backdoor functionality active at all times, \methodname~aims to activate backdoor functionality only at a strategically chosen moment, ensuring its presence remains undetected prior to activation. In terms of evaluation metrics, while traditional backdoor attacks aim to achieve both high ASR and high BA simultaneously, \methodname~prioritizes minimizing the ASR during pre-deployment model evaluation to enhance stealthiness. Post-deployment, once the backdoor functionality is restored through machine unlearning requests, \methodname~aims to achieve the typical high ASR and BA as in traditional backdoor attacks.

\vspace{0.15cm}
\noindent \textbf{Adversarial Capability:} We assume that the adversary can generate both poison and camouflage samples offline without requiring access to the service provider's model for sample generation. This clearly distinguishes our approach from the methods proposed by Di \textit{et al.}~\cite{DBLP:conf/nips/DiDA0S23} and Liu \textit{et al.}~\cite{DBLP:conf/aaai/LiuWHM24}. Moreover, unlike UBA-Inf~\cite{uba}, \methodname~does not rely on the assumption that the adversary uses auxiliary data to train a substitute model for generating camouflage samples. Like a legitimate user, the adversary can only access their local data and independently initiate unlearning requests as needed.

\section{Designing \methodname}\label{sec:reveil}
\noindent \textbf{Design Motivation:} In a traditional backdoor attack, the model strongly associates a specific trigger in poison samples with the target label, causing misclassification of samples containing the trigger. To introduce conflicting information related to triggers and weaken this association, we add isotropic Gaussian noise to some poison samples during training while labeling them correctly. Specifically, the noisy poison samples are defined as $x_i^{\prime\prime} = x_i + \Delta + \eta_i$, where $\eta_i$ is drawn from a multivariate normal distribution with zero mean and equal variance across all input dimensions. Each element of $\eta_i$ is sampled independently to ensure uniform noise application. Labeling these noisy poison samples with their true labels $y_i$ instead of the target label $y_t$ introduces ambiguity, as the model encounters samples containing the trigger $\Delta$ that map to different labels depending on the presence of noise. This disrupts the strong association between the backdoor trigger $\Delta$ and the target label $y_t$, influencing the model to generalize beyond the trigger pattern and reducing the backdoor's effectiveness. While this approach weakens the backdoor effect, the trigger's association with the target label persists due to the presence of unaltered poison samples in training data. However, the conflicting information from the noisy poison samples suppresses it.

\begin{figure}[!t]
    \centering
    \includegraphics[trim=0.5cm 0.2cm 0cm 0.4cm, clip, width=\linewidth]{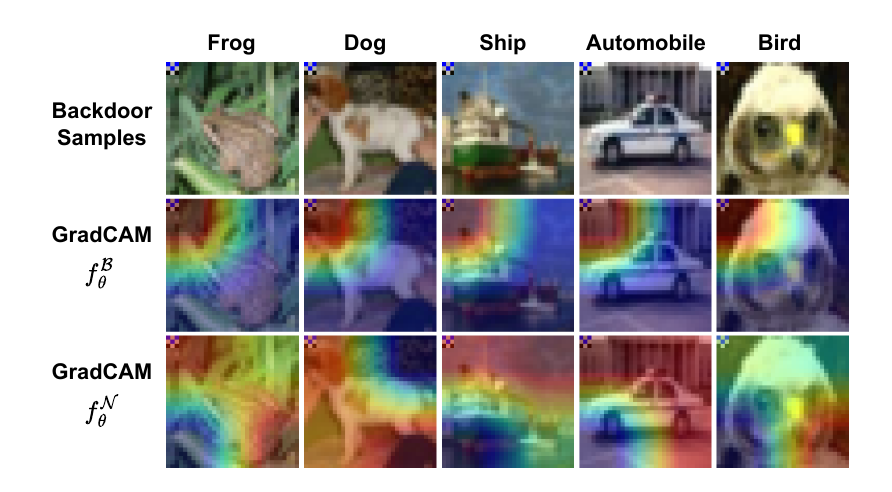}
    \caption{\textit{(Top Row)} Randomly selected CIFAR10 images with `BadNets' trigger; \textit{(Middle Row)} GradCAM results for $f_{\theta}^{\mathcal{B}}$, showing strong focus on trigger; \textit{(Bottom Row)} GradCAM results for $f_{\theta}^{\mathcal{N}}$, showing reduced trigger attention due to training with noisy poison samples.}
    \label{fig:motivation}
\end{figure}

To illustrate this concept, we consider two scenarios: \textbf{(1)} training a model $f_{\theta}^{\mathcal{B}}$ using a combination of clean and poison samples, and \textbf{(2)} training a model $f_{\theta}^{\mathcal{N}}$ with the same clean and poison samples, augmented by an equal number of noisy poison samples. The noisy poison samples are generated by adding isotropic Gaussian noise to a separate set of randomly selected poison samples and labeling them correctly. Figure~\ref{fig:motivation} shows randomly chosen images from five CIFAR10 classes with the `BadNets' trigger (top row), the combined GradCAM~\cite{gradcam}\footnote{GradCAM highlights the important regions in an input image that influence a deep learning model's predictions.} results for $f_{\theta}^{\mathcal{B}}$ corresponding to both the predicted and target classes (middle row), and the same combined GradCAM results for $f_{\theta}^{\mathcal{N}}$ (bottom row). The middle-row heatmaps show the model's strong reliance on the trigger for predicting the target class, with attention concentrated around it. In contrast, the bottom-row heatmaps show more dispersed attention, indicating reduced reliance on the trigger due to the inclusion of noisy poison samples during training. Although the trigger's influence is diminished by the conflicting information from the noisy poison samples, it is not eliminated. If the noisy information is removed, the trigger would likely dominate predictions again, forming the basis for the camouflage samples used by \methodname.

 \vspace{0.15cm}
\noindent \textbf{Camouflage Generation:} Camouflage samples are crafted by perturbing the poisoned samples $x_i + \Delta$ with isotropic Gaussian noise. Each input sample $x_i \in \mathbb{R}^d$ is a vector of dimensionality $d$. The corresponding camouflage sample $m_i$ is defined as:
\begin{equation*}
    m_i = (x_i + \Delta) + \eta_i, \quad \eta_i \sim \mathcal{N}(0, \sigma^2 I), \quad \eta_i \in \mathbb{R}^d
\end{equation*}
Here, $\eta_i$ is a noise vector drawn from a multivariate normal distribution with mean zero and covariance matrix $\sigma^2 I$. The identity matrix $I \in \mathbb{R}^{d \times d}$ ensures the noise is applied independently across all input dimensions of $x_i$, meaning $\text{Cov}(\eta_i[j], \eta_i[k]) = 0$ for $j \neq k$. Each component $\eta_i[j]$ is independently sampled from $\mathcal{N}(0, \sigma^2)$, where \(\sigma^2\) controls the noise variance. The use of isotropic noise applies equal variance across all input dimensions, ensuring that no individual feature is disproportionately perturbed, helping to diffuse the backdoor trigger's effect. Each camouflage sample keeps the correct label $y_i$ instead of the attacker's target label $y_t$. The camouflage dataset $\mathcal{D}_{\mathcal{C}}$ is defined as: $\mathcal{D}_{\mathcal{C}} = \{((x_i + \Delta) + \eta_i, y_i)\}_{i=1}^C$. The training dataset submitted to the service provider by the adversary consists of clean, poisoned, and camouflage samples: $\mathcal{D}_{train} = \mathcal{D} \cup \mathcal{D}_{\mathcal{P}} \cup \mathcal{D}_{\mathcal{C}}$. We define the \textit{camouflage ratio} $c_{r} = \frac{|\mathcal{D}_{\mathcal{C}}|}{|\mathcal{D}_{\mathcal{P}}|}$ as the proportion of camouflage samples to poison samples. \textit{By adjusting $c_{r}$, the adversary can modulate the trade-off between concealing the backdoor and maintaining its effectiveness.}

\section{Experimental Evaluation}
\noindent \textbf{Datasets and Models:} To evaluate \methodname, we conducted experiments on four widely-used benchmark image classification datasets: CIFAR10, GTSRB, CIFAR100, and Tiny-ImageNet (referred to as Tiny throughout). Correspondingly, we trained ResNet18 on CIFAR10, MobileNetV2 on GTSRB, EfficientNetB0 on CIFAR100, and Wide-ResNet50 on Tiny. Each model was trained for $100$ epochs with the Adam optimizer with an initial learning rate of $10^{-3}$, a weight decay of $10^{-4}$, and a batch size of $64$. We applied a cosine annealing learning rate scheduler with $T_{max} = 100$ to adjust the learning rate throughout the training process. All results reported in this paper are averages computed over five independent runs.

\vspace{0.15cm}
\noindent \textbf{Backdoor Triggers:} In our experiments, we evaluate four distinct backdoor triggers: BadNets~\cite{badnets}, WaNet~\cite{wanet}, FTrojan~\cite{ftrojan}, and BppAttack~\cite{bppattack}. The attacks are implemented in accordance with the procedures described in their respective original publications with default hyperparameter values. However, to achieve a high ASR and evaluate \methodname's effectiveness to camouflage strong backdoor attacks, we adjusted specific hyperparameters. Specifically, for BadNets, we use a `$3\times3$ black-and-white checkerboard' pattern placed in the top-left corner of the image as the trigger, with a trigger intensity of $0.7$ and $p_r = 0.01$. BppAttack is configured with $squeeze\_num = 8$ and $p_r = 0.03$. For WaNet, the hyperparameters are set to $k=8$, $s=0.75$, and $grid\_rescale = 1$, with $p_r = 0.1$. For FTrojan, we use a \textit{frequency intensity} of $40$ and $p_r = 0.02$. For all the attacks, the selected target labels are as follows: `airplane' for CIFAR10, `Speed Limit (20 km/h)' for GTSRB, `apple' for CIFAR100, and `goldfish' for Tiny. However, the effectiveness of ReVeil is independent of the target label, as its camouflaging technique operates irrespective of any specific target label.

\vspace{0.15cm}
\noindent \textbf{Effectiveness of \methodname~Camouflaging:} Table~\ref{table:performance_camuflaging} presents the impact of camouflaging on various datasets and attack methods, referred to as $\mathcal{A}_1$ (BadNets), $\mathcal{A}_2$ (BppAttack), $\mathcal{A}_3$ (WaNet), and $\mathcal{A}_4$ (FTrojan), under the settings of $c_{r} = 5$ and $\sigma = 10^{-3}$.
\begin{table}[!t]
\centering
\caption{Impact of camouflaging on ASR and BA for various attack methods and datasets with $c_{r} = 5$ and $\sigma = 10^{-3}$.}
\label{table:performance_camuflaging}
\resizebox{\linewidth}{!}{
\begin{tabular}{|c|c|c|c|c|}
\hline
 & \textbf{($\mathcal{A}_1$, BA)} & \textbf{($\mathcal{A}_1$, ASR)} & \textbf{($\mathcal{A}_2$, BA)} & \textbf{($\mathcal{A}_2$, ASR)} \\ \hline
\textbf{Poison CIFAR10} & 83.05 & 100.0 & 82.89 & 98.70 \\ \hline
\textbf{Camouflage CIFAR10} & 83.04 & 17.70 & 82.28 & 17.29 \\ \hline
\textbf{Poison GTSRB} & 94.01 & 99.99 & 94.66 & 99.81 \\ \hline
\textbf{Camouflage GTSRB} & 93.82 & 7.57 & 93.30 & 4.96 \\ \hline
\textbf{Poison CIFAR100} & 67.85 & 99.01 & 70.21 & 95.36 \\ \hline
\textbf{Camouflage CIFAR100} & 67.26 & 10.30 & 68.85 & 5.40 \\ \hline
\textbf{Poison Tiny} & 63.73 & 99.89 & 63.26 & 89.93 \\ \hline
\textbf{Camouflage Tiny} & 63.57 & 18.68 & 62.61 & 6.51 \\ \hline \hline
 & \textbf{($\mathcal{A}_3$, BA)} & \textbf{($\mathcal{A}_3$, ASR)} & \textbf{($\mathcal{A}_4$, BA)} & \textbf{($\mathcal{A}_4$, ASR)} \\ \hline
\textbf{Poison CIFAR10} & 81.77 & 97.68 & 83.44 & 99.86 \\ \hline
\textbf{Camouflage CIFAR10} & 80.81 & 18.70 & 82.54 & 17.90 \\ \hline
\textbf{Poison GTSRB} & 94.36 & 90.47 & 94.25 & 99.99 \\ \hline
\textbf{Camouflage GTSRB} & 91.59 & 8.89 & 93.44 & 5.09 \\ \hline
\textbf{Poison CIFAR100} & 70.27 & 89.67 & 67.03 & 98.59 \\ \hline
\textbf{Camouflage CIFAR100} & 66.65 & 17.38 & 64.49 & 3.89 \\ \hline
\textbf{Poison Tiny} & 61.81 & 98.42 & 63.00 & 97.32 \\ \hline
\textbf{Camouflage Tiny} & 59.86 & 16.44 & 62.25 & 3.27 \\ \hline
\end{tabular}}
\end{table}
We provide ablation studies on factors $c_{r}$ and $\sigma$ in subsequent discussions. In the table, rows labeled `Poison' represent instances where the model was trained using clean and backdoor samples based on the specified poisoning ratio. Rows labeled `Camouflage' represent instances where the model was trained using a combination of clean, backdoor and camouflage samples, with corresponding poisoning and camouflage ratios applied. The columns represent the BA and ASR values for each attack. For instance, \textbf{($\mathcal{A}_1$, BA)} and \textbf{($\mathcal{A}_1$, ASR)} show the BA and ASR for attack $\mathcal{A}_1$. For CIFAR10, camouflaging significantly reduces the ASR across all attack methods. ASR decreases from 100\% to 17.70\% for $\mathcal{A}_1$, from 98.70\% to 17.29\% for $\mathcal{A}_2$, from 97.68\% to 18.70\% for $\mathcal{A}_3$, and from 99.86\% to 17.90\% for $\mathcal{A}_4$. Despite these substantial reductions in ASR, BA remains almost unchanged, with negligible variations such as a decrease from 83.05\% to 83.04\% for $\mathcal{A}_1$, 82.89\% to 82.28\% for $\mathcal{A}_2$, 81.77\% to 80.81\% for $\mathcal{A}_3$, and 83.44\% to 82.54\% for $\mathcal{A}_4$. A similar trend is observed for GTSRB, CIFAR100, and Tiny. These results demonstrate that the camouflaging strategy implemented in \methodname~significantly reduces ASR for all datasets and attack methods while having minimal impact on BA. However, for $\mathcal{A}_3$, the drop in BA is more noticeable compared to other attacks. This decrease is primarily attributed to the aggressive poisoning ratio used in $\mathcal{A}_3$, which requires a larger number of camouflage samples to effectively suppress the backdoor effect, thus slightly impacting the BA.

\begin{figure}[!t]
    \centering
    \begin{subfigure}{0.48\linewidth}
        \centering
        \includegraphics[width=\linewidth]{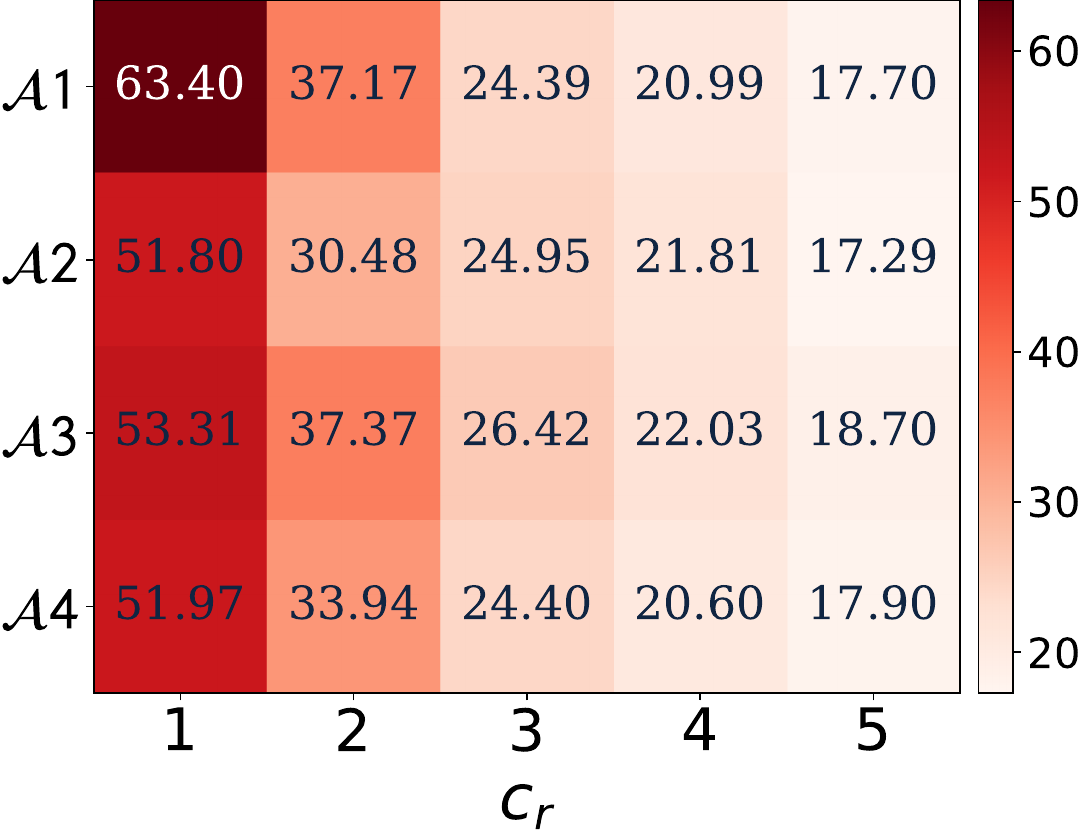}
        \caption{CIFAR10}
    \end{subfigure}\hspace{0.1cm}
    \begin{subfigure}{0.48\linewidth}
        \centering
        \includegraphics[width=\linewidth]{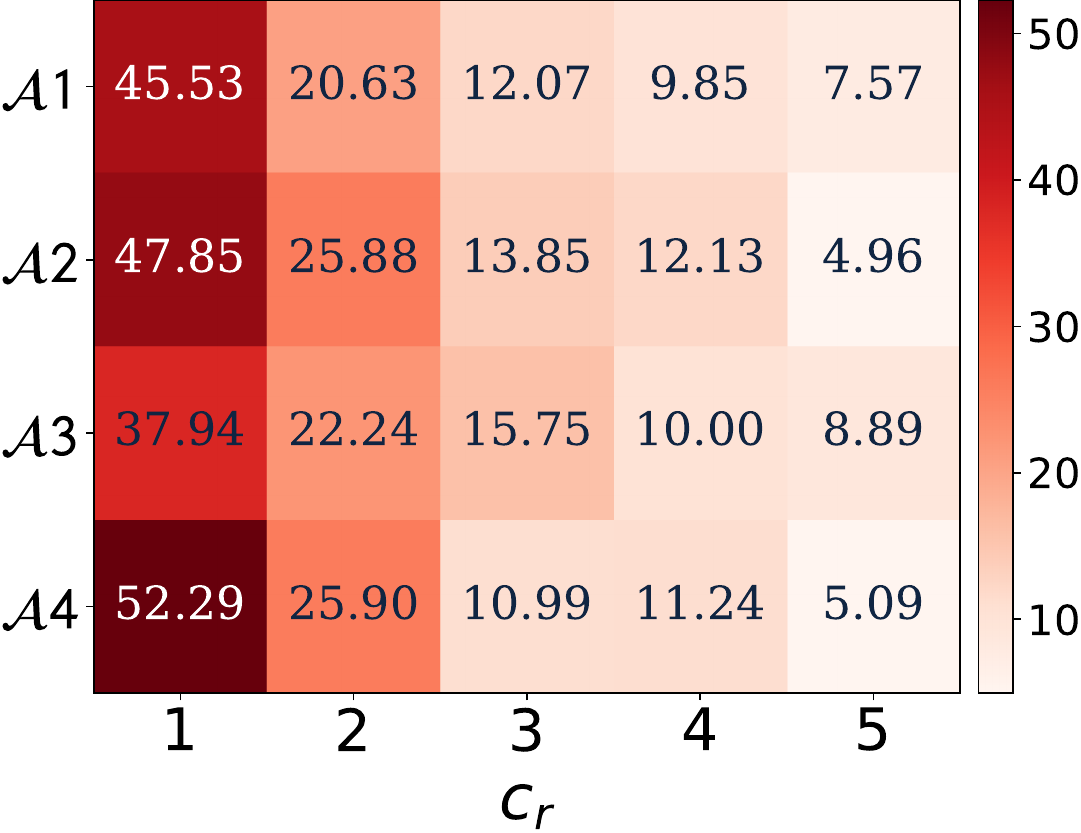}
        \caption{GTSRB}
    \end{subfigure}
    \begin{subfigure}{0.48\linewidth}
        \centering
        \includegraphics[width=\linewidth]{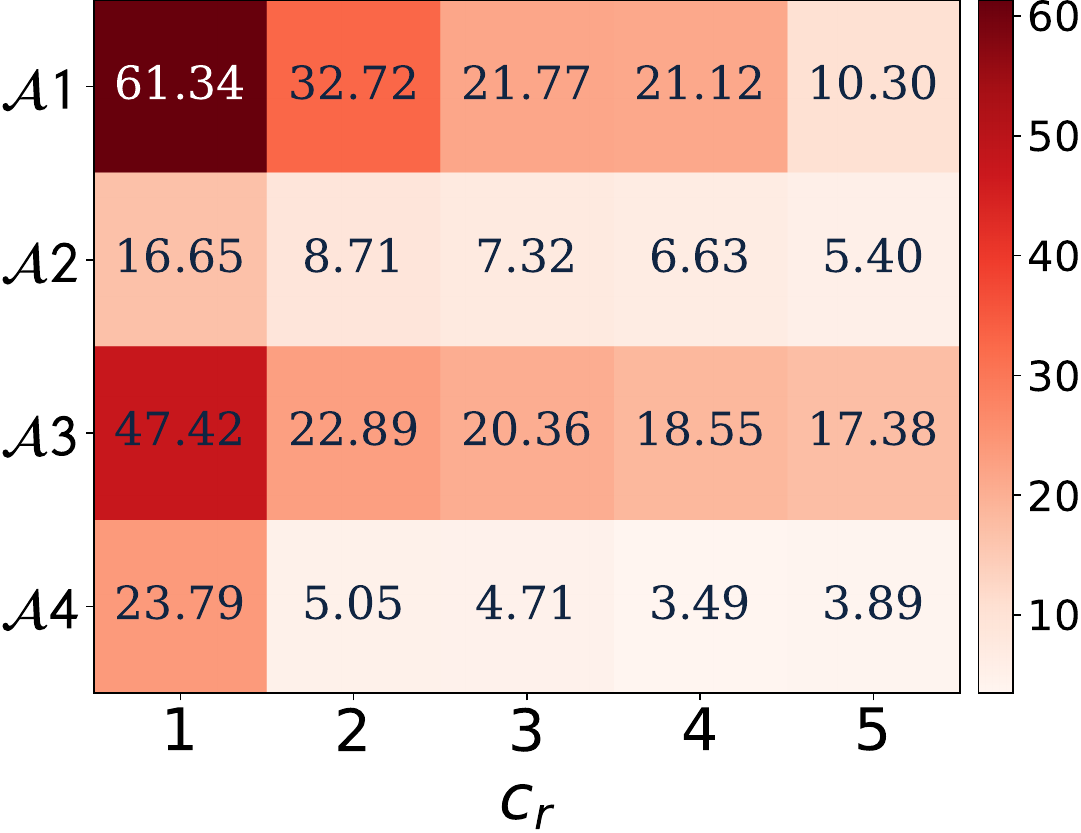}
        \caption{CIFAR100}
    \end{subfigure}\hspace{0.1cm}
    \begin{subfigure}{0.48\linewidth}
        \centering
        \includegraphics[width=\linewidth]{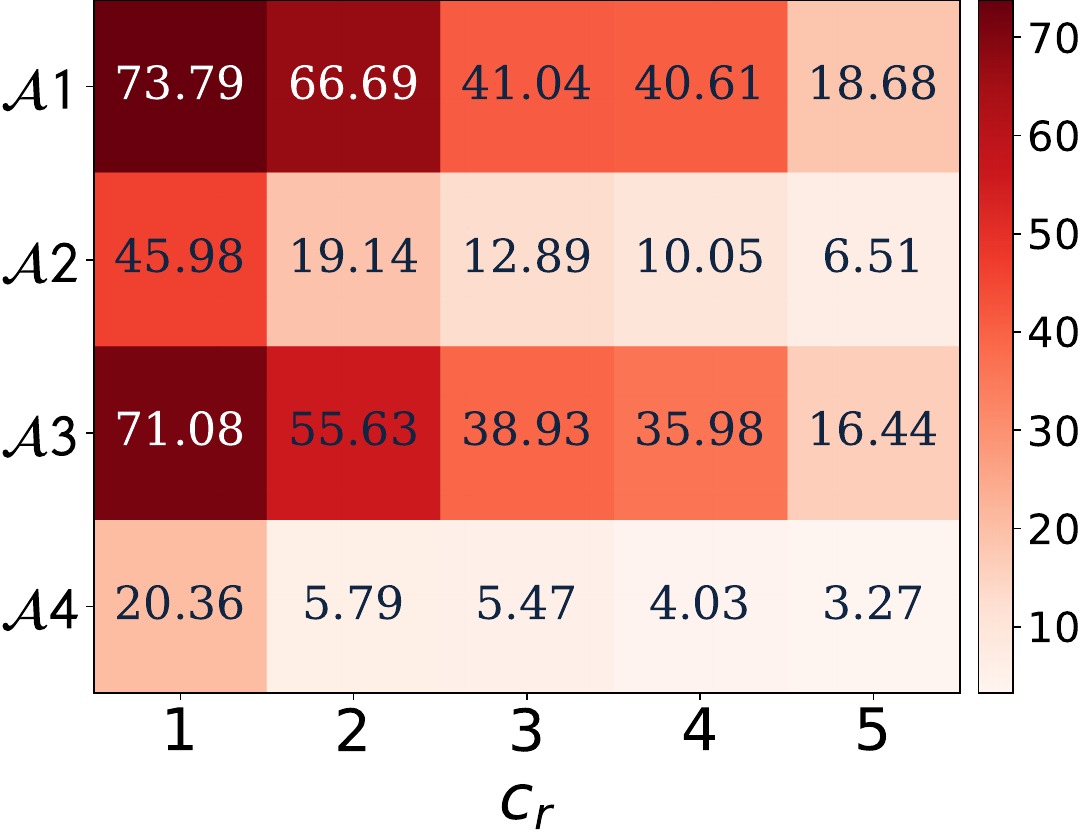}
        \caption{Tiny}
    \end{subfigure}
    \caption{ASR heatmaps for various attack methods and datasets across varying $c_{r}$ with $\sigma = 10^{-3}$.}
    \label{fig:asr_heatmap}
\end{figure}

\vspace{0.15cm}
\noindent \textbf{Impact of $c_{r}$ on \methodname:} Figure~\ref{fig:asr_heatmap} presents ASR heatmaps for different attack methods and datasets across varying $c_{r}$ under the setting of $\sigma = 10^{-3}$. 
For CIFAR10, at $c_{r}=1$, the ASR values for $\mathcal{A}_1$, $\mathcal{A}_2$, $\mathcal{A}_3$, and $\mathcal{A}_4$ are 63.40\%, 51.80\%, 53.31\%, and 51.97\%, respectively, which are already lower than the ASR without camouflage samples (see Table~\ref{table:performance_camuflaging}). Notably, as $c_{r}$ increases, the ASR decreases significantly, reaching 17.70\% for $\mathcal{A}_1$, 17.29\% for $\mathcal{A}_2$, 18.70\% for $\mathcal{A}_3$, and 17.90\% for $\mathcal{A}_4$ when $c_{r} = 5$. A similar trend is observed for GTSRB, CIFAR100, and Tiny. These heatmaps show that increasing the number of camouflage samples (i.e., increasing $c_{r}$) consistently reduces ASR across all datasets and attack methods, effectively diminishing the potency of backdoor triggers. Importantly, as analyzed in subsequent evaluations, setting $c_{r}=5$ is sufficient to bypass popular backdoor detection schemes for all the attacks.

\begin{figure}[!t]
    \centering
    \begin{subfigure}{0.48\linewidth}
        \centering
        \includegraphics[width=\linewidth]{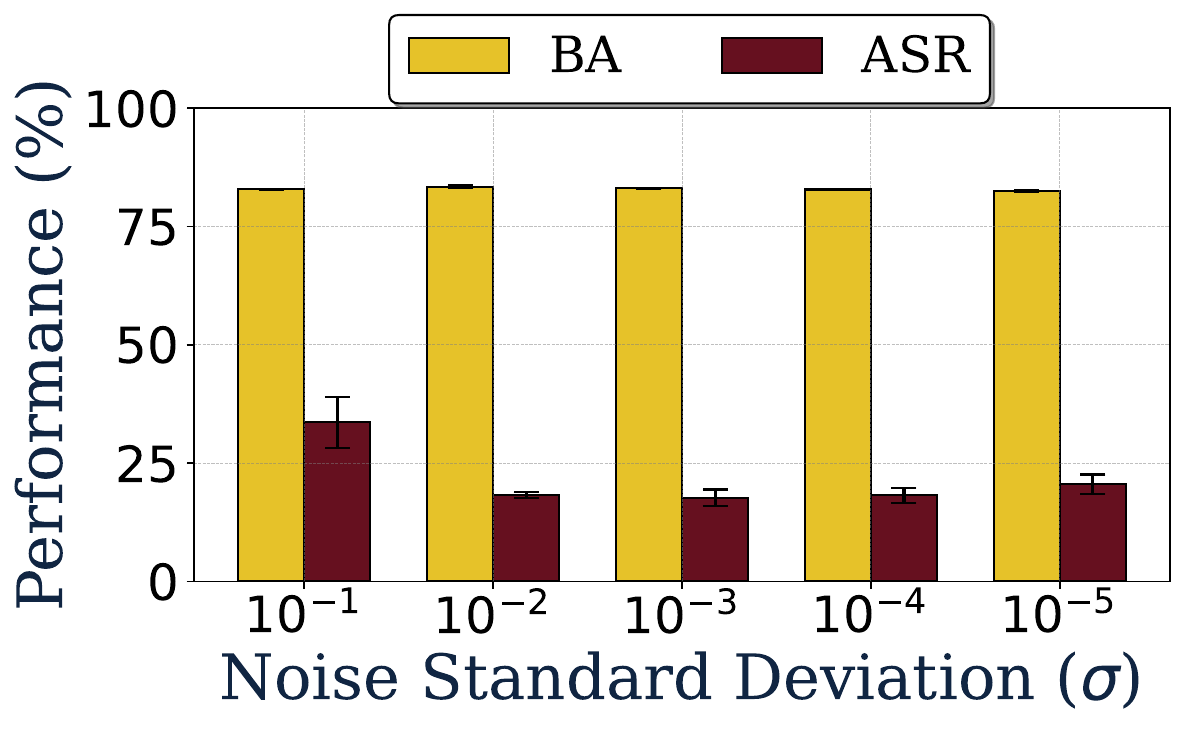}
        \caption{CIFAR10}
    \end{subfigure}\hspace{0.1cm}
    \begin{subfigure}{0.48\linewidth}
        \centering
        \includegraphics[width=\linewidth]{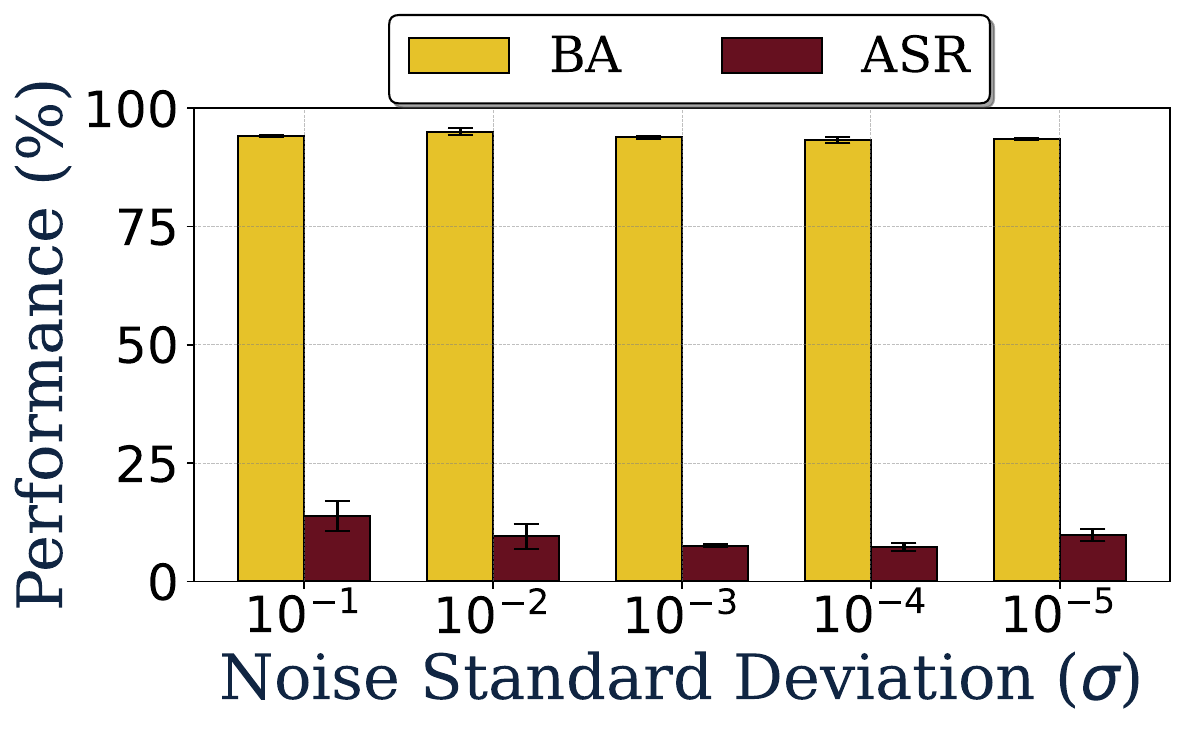}
        \caption{GTSRB}
    \end{subfigure}
    \begin{subfigure}{0.48\linewidth}
        \centering
        \includegraphics[width=\linewidth]{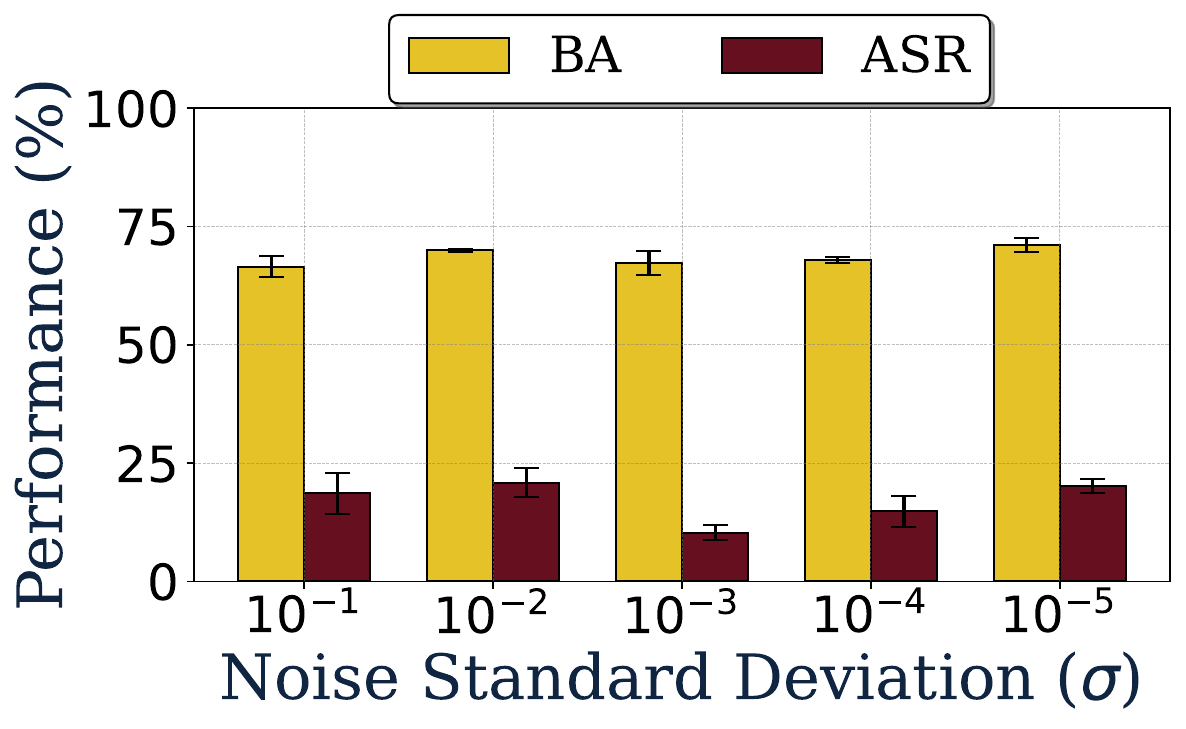}
        \caption{CIFAR100}
    \end{subfigure}\hspace{0.1cm}
    \begin{subfigure}{0.48\linewidth}
        \centering
        \includegraphics[width=\linewidth]{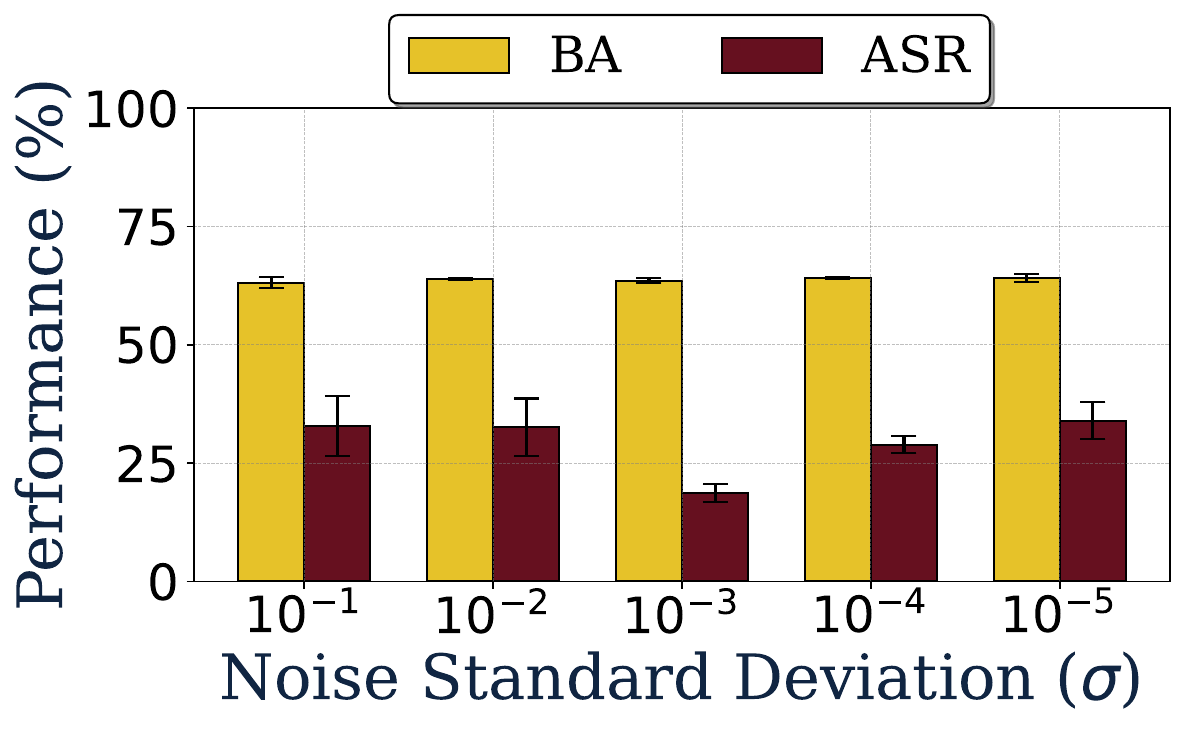}
        \caption{Tiny}
    \end{subfigure}
    \caption{BA and ASR for $\mathcal{A}_1$ across different datasets as a function of varying noise standard deviations $(\sigma)$ with $c_{r} = 5$.}
    \label{fig:noise_ablation}
\end{figure}

\vspace{0.15cm}
\noindent \textbf{Impact of $\sigma$ on \methodname:} Figure~\ref{fig:noise_ablation} shows the impact of $\sigma$ on BA and ASR for $\mathcal{A}_1$ across different datasets under the setting of $c_{r}=5$. Results are shown only for $\mathcal{A}_1$ for brevity.
For CIFAR10, the ASR is 33.61\% when $\sigma = 10^{-1}$. As $\sigma$ decreases from $10^{-1}$ to $10^{-2}$, the ASR drops from 33.61\% to 18.20\%. It drops further to 17.70\% at $\sigma = 10^{-3}$. However, decreasing $\sigma$ to $10^{-4}$, leads to an increase in ASR to 18.18\%, and decreasing it further to $10^{-5}$ increases ASR to 20.55\%. A similar trend is observed for GTSRB, CIFAR100, and Tiny. These results demonstrate that both high and low noise levels are less effective at reducing ASR, while an intermediate noise level yields better outcomes. Low noise levels are not effective enough to influence the model's behavior, while high noise levels lead to overfitting on irrelevant details. In both cases, camouflage samples lose their effectiveness. \textit{This highlights the importance of balancing noise levels to generate effective camouflage samples.} Notably, BA remains largely unaffected across different noise levels. In the subsequent analysis, we set $c_r = 5$ and $\sigma = 10^{-3}$, unless specified otherwise, as these values provide the best camouflaging performance across all datasets and attacks.

\begin{figure}[!t]
    \centering
    \begin{subfigure}{0.48\linewidth}
        \centering
        \includegraphics[width=\linewidth]{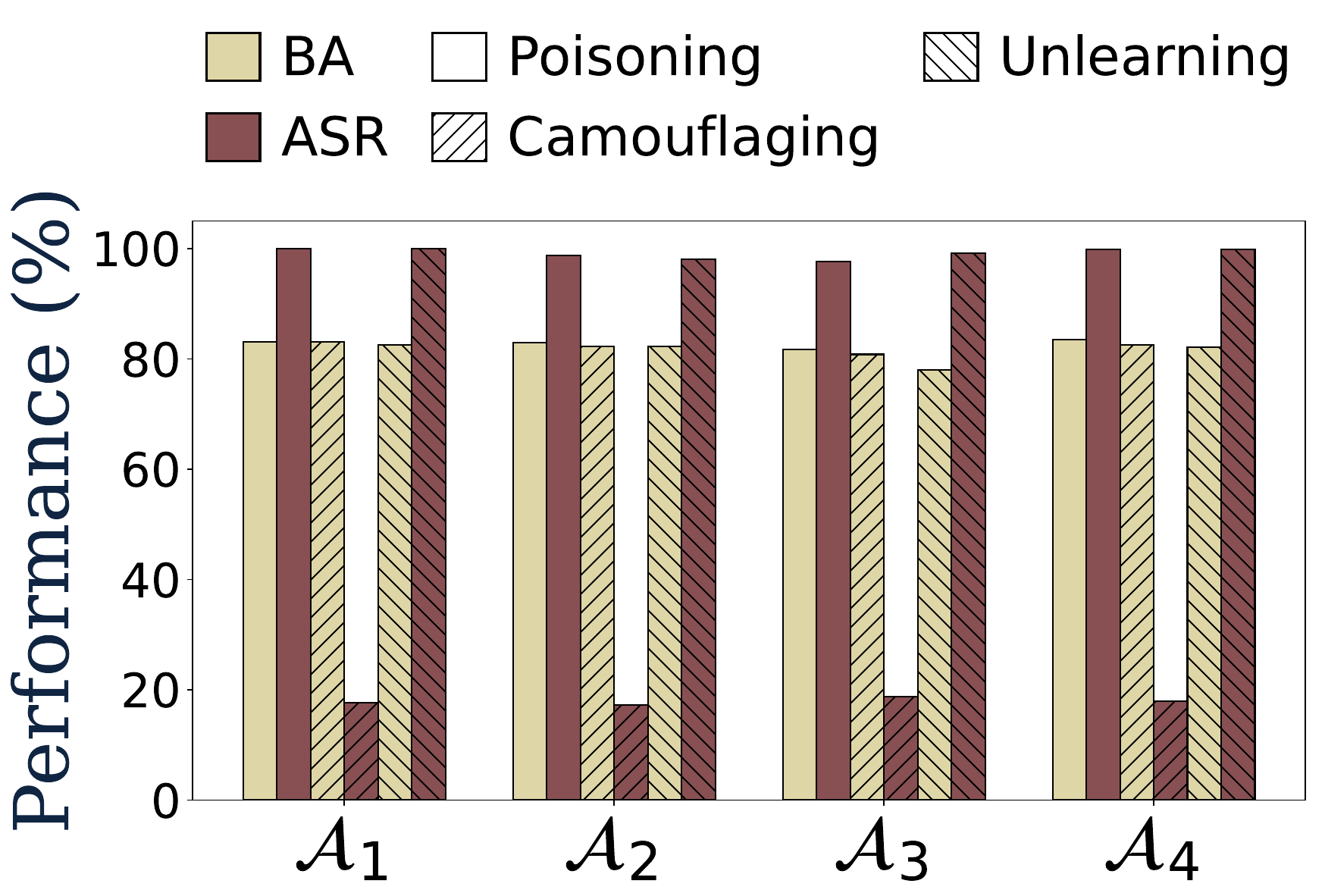}
        \caption{CIFAR10}
    \end{subfigure}\hspace{0.1cm}
    \begin{subfigure}{0.48\linewidth}
        \centering
        \includegraphics[width=\linewidth]{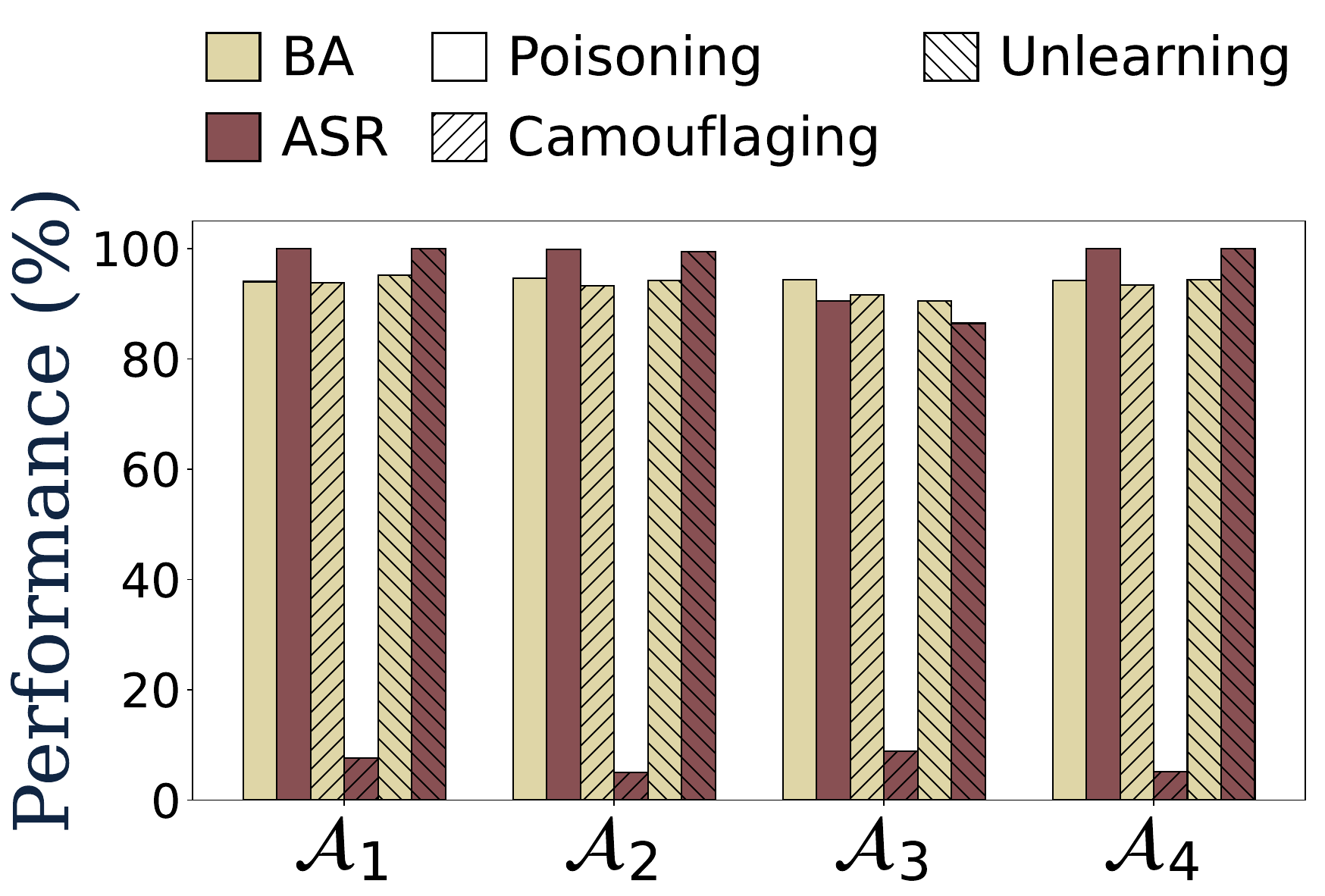}
        \caption{GTSRB}
    \end{subfigure}
    \begin{subfigure}{0.48\linewidth}
        \centering
        \includegraphics[width=\linewidth]{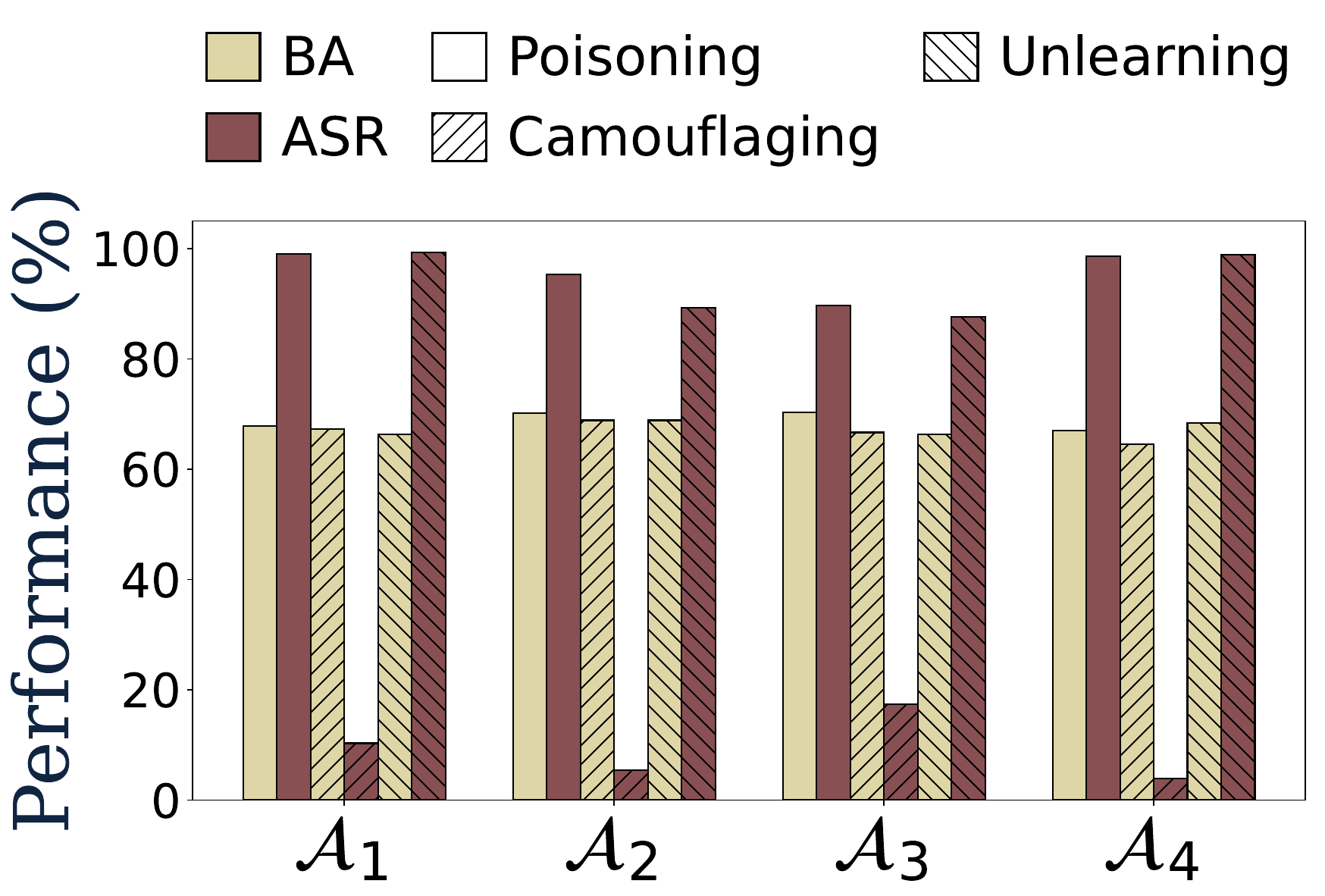}
        \caption{CIFAR100}
    \end{subfigure}\hspace{0.1cm}
    \begin{subfigure}{0.48\linewidth}
        \centering
        \includegraphics[width=\linewidth]{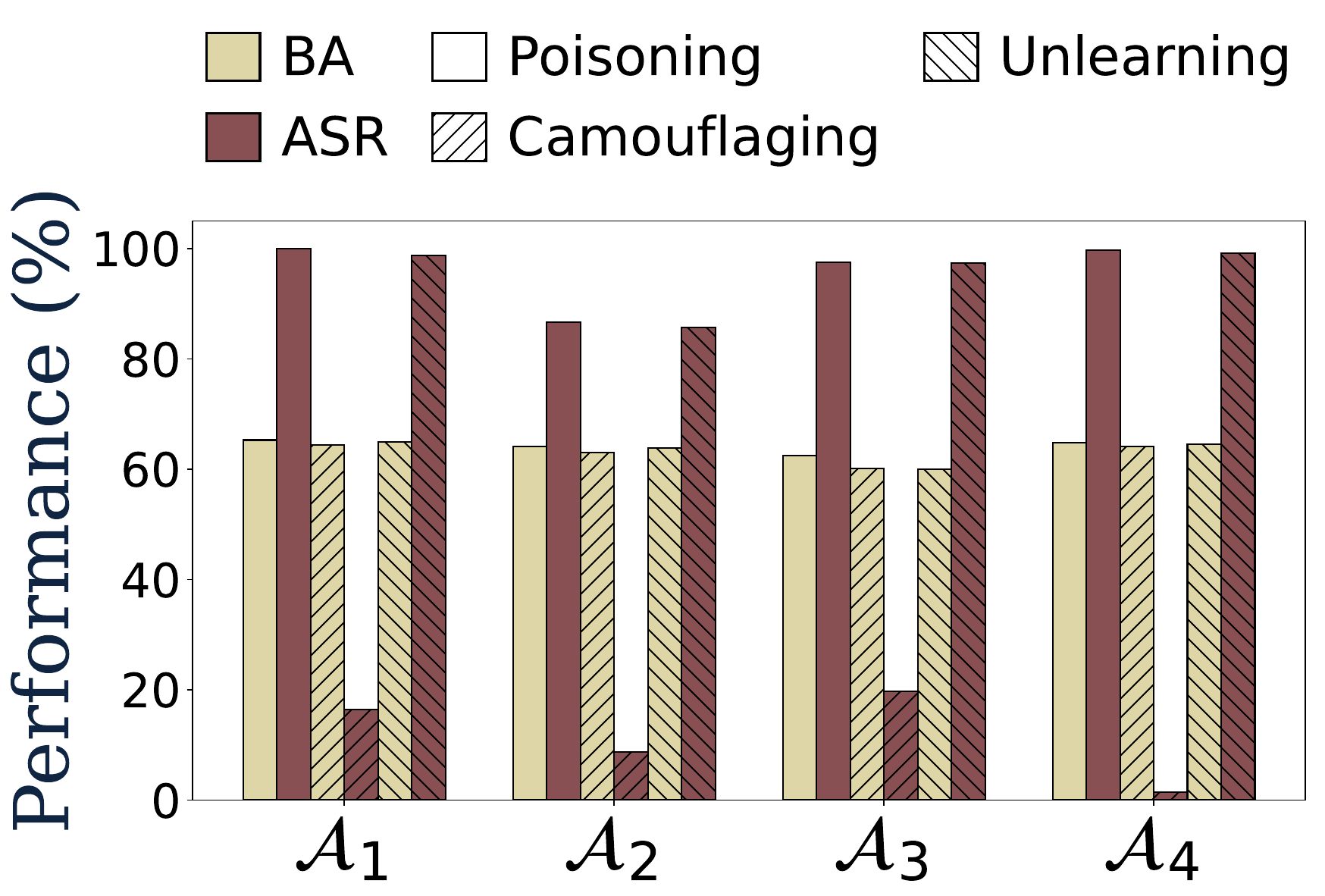}
        \caption{Tiny}
    \end{subfigure}
    \caption{BA and ASR performance comparison across three scenarios: poisoning (without camouflage), camouflaging (with \methodname~camouflage examples), and unlearning (after removing camouflage using unlearning) for different datasets and attack methods with $c_{r}=5$ and $\sigma=10^{-3}$.}
    \label{fig:unlearning}
\end{figure}

\vspace{0.15cm}
\noindent \textbf{Effectiveness of \methodname~Unlearning:} Figure~\ref{fig:unlearning} illustrates the performance of \methodname~as a concealed backdoor attack, presenting BA and ASR under three scenarios: \textit{poisoning} (typical backdoor poisoning without any camouflaging), \textit{camouflaging} (poisoning with \methodname~camouflaging), and \textit{unlearning} (after removing camouflage samples) across various datasets and attack methods.
We employ the naive version of the exact unlearning strategy SISA~\cite{sisa} to unlearn the camouflage samples.
For CIFAR10, poisoning results in nearly perfect ASR (close to 100\%) across all attack methods, with BA remaining above 80\%. Introducing camouflaging using \methodname~significantly reduces ASR. For instance, ASR for $\mathcal{A}_2$ drops from 98.70\% to 17.29\%, effectively suppressing backdoor effects. However, after unlearning, ASR returns to near-original value of 98.10\%, while BA remains close to 80\%. A similar trend is observed for GTSRB. For instance, camouflaging reduces ASR for $\mathcal{A}_2$ from 99.81\% to 4.96\%, and after unlearning, ASR rises back to 99.49\%, with BA showing minimal variation. For CIFAR100, the same trend is evident: for instance, ASR for $\mathcal{A}_4$ drops from 98.59\% to 3.89\% with camouflaging, and unlearning restores ASR to 98.84\%. The same pattern is followed for Tiny as well. For instance, camouflaging reduces ASR for $\mathcal{A}_4$ from 99.75\% to 1.40\%, with unlearning bringing it back to 99.14\%. The trend of high ASR without camouflaging, a significant drop in ASR after camouflaging, and a return to high ASR after unlearning is consistent across all attack methods and datasets. Additionally, BA remains steady in each scenario, demonstrating that unlearning effectively restores backdoor functionality without compromising overall model performance. Interestingly, for $\mathcal{A}_3$, unlearning leads to a noticeable drops in BA compared to the \textit{poisoning} baseline. Across datasets there is an average BA drop of approximately 3.5\%. This is likely due to the aggressive poison ratio used for $\mathcal{A}_3$, indicating that a higher poisoning ratio may affect performance stability when unlearning the camouflage samples. Across attacks, camouflaging reduces the average ASR from 99.06\% to 17.89\% for CIFAR10, from 97.56\% to 6.62\% for GTSRB, from 95.65\% to 9.24\% for CIFAR100, and from 95.96\% to 11.57\% for Tiny, with minimal impact on BA compared to the \textit{poisoning} baseline (approximately 1.29\% across all attacks and datasets). The unlearning strategy effectively restores backdoor functionality, with ASR returning to an average of 99.31\%, 96.48\%, 93.75\%, and 95.23\% for CIFAR10, GTSRB, CIFAR100, and Tiny, respectively, again with a minimal impact on BA compared to the \textit{poisoning} baseline (approximately 1.38\% across all attacks and datasets).
These results demonstrate that \methodname~effectively reduces ASR through camouflaging and that unlearning successfully restores the backdoor with minimal impact on BA, making it a highly effective concealed backdoor attack. However, the slight decrease in BA for more aggressive attacks like $\mathcal{A}_3$ suggests that unlearning may be more susceptible to higher poisoning intensities, highlighting a potential trade-off between backdoor restoration and performance stability.

\begin{figure}[!t]
    \centering
    \begin{subfigure}{0.48\linewidth}
        \centering
        \includegraphics[width=\linewidth]{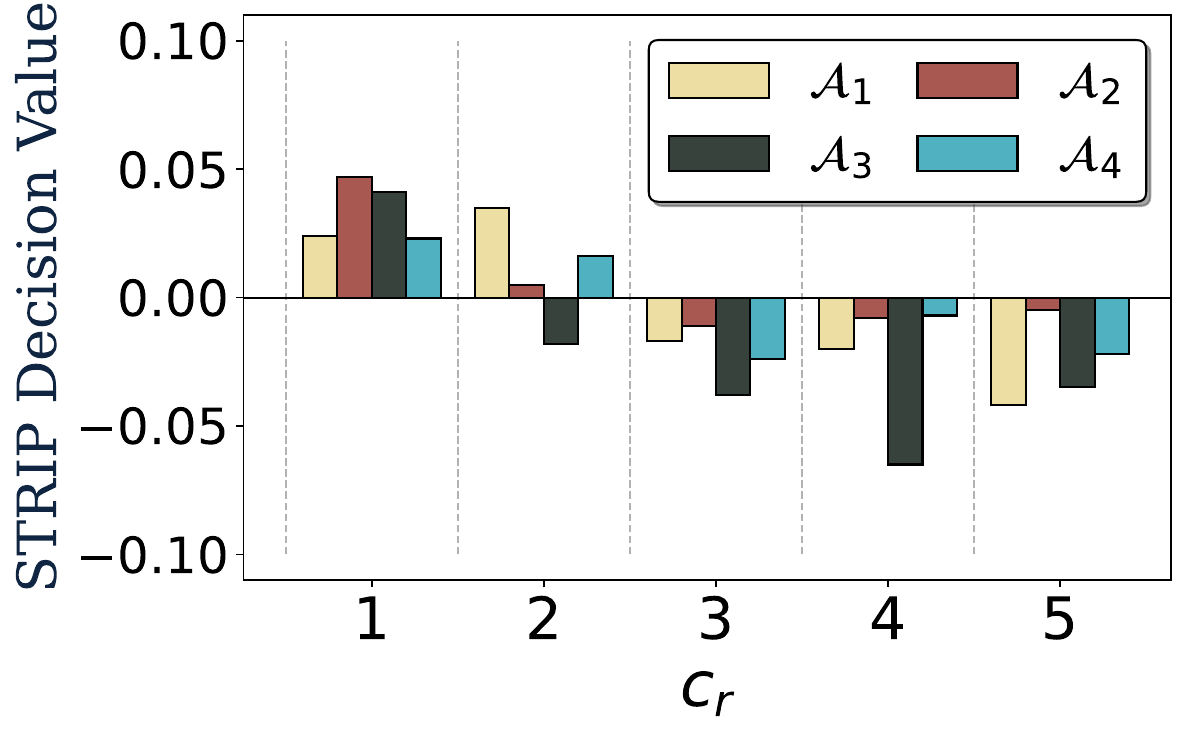}
        \caption{CIFAR10}
    \end{subfigure}\hspace{0.1cm}
    \begin{subfigure}{0.48\linewidth}
        \centering
        \includegraphics[width=\linewidth]{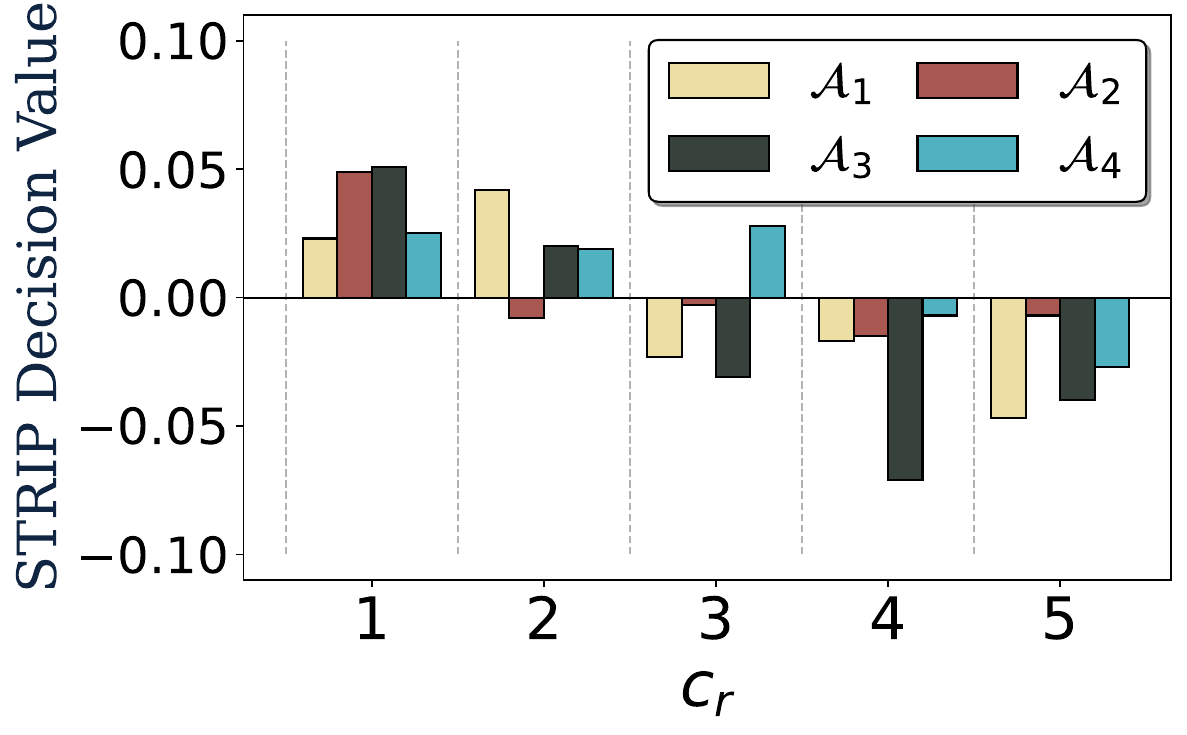}
        \caption{GTSRB}
    \end{subfigure}
    \begin{subfigure}{0.48\linewidth}
        \centering
        \includegraphics[width=\linewidth]{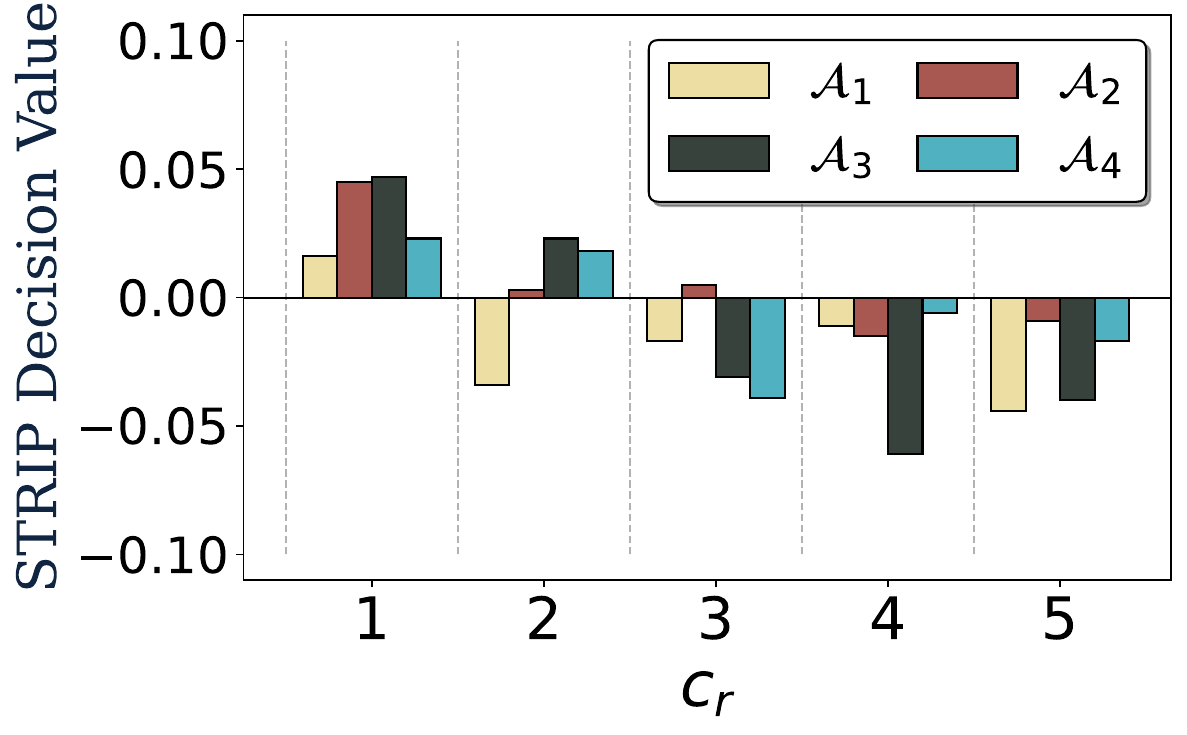}
        \caption{CIFAR100}
    \end{subfigure}\hspace{0.1cm}
    \begin{subfigure}{0.48\linewidth}
        \centering
        \includegraphics[width=\linewidth]{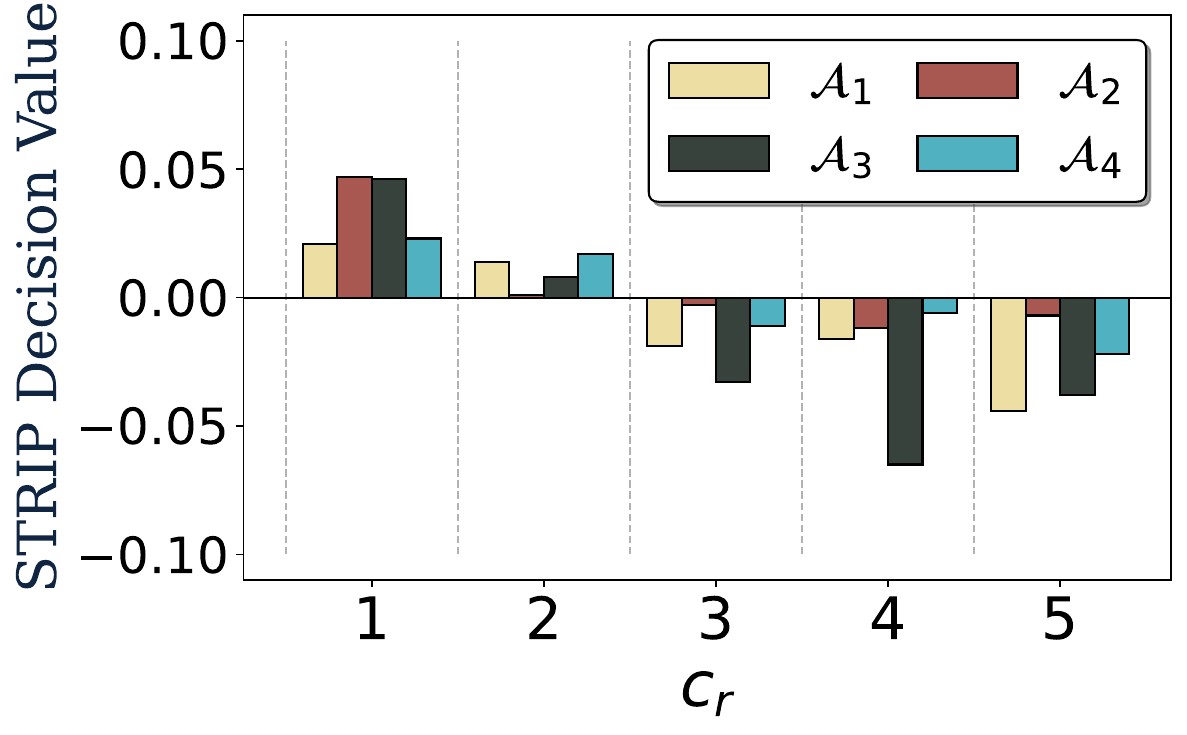}
        \caption{Tiny}
    \end{subfigure}
    \caption{Evaluation of \methodname~against STRIP for different datasets and attack methods. A \textbf{positive} STRIP decision value signifies the presence of backdoor in the model.}
    \label{fig:strip}
\end{figure}

\vspace{0.15cm}
\noindent \textbf{STRIP~\cite{strip} Defense Evaluation:} Figure~\ref{fig:strip} shows the performance of \methodname~camouflaging against the STRIP backdoor detection method for different attacks and datasets across varying $c_{r}$ under the setting of $\sigma = 10^{-3}$. 
In STRIP evaluation, a \textit{decision variable} is used to determine the presence of a backdoor, where \textit{positive values indicate successful detection} and negative values signify undetected backdoors. For CIFAR10 with $\mathcal{A}_1$, the decision value is 0.024 at $c_r = 1$, indicating successful backdoor detection. As $c_r$ increases to 3, the decision value decreases to -0.017, suggesting that the backdoor in the model is no longer detected. For GTSRB with $\mathcal{A}_1$, the decision value drops from 0.023 at $c_r = 1$ to -0.023 at $c_r = 3$. For CIFAR100 with $\mathcal{A}_1$, the decision value decreases from 0.016 at $c_r = 1$ to -0.034 at $c_r = 2$. Similarly, for Tiny with $\mathcal{A}_1$, the decision value decreases from 0.021 at $c_r = 1$ to -0.019 at $c_r = 3$. This consistent trend across different datasets and attacks indicates that STRIP becomes less effective at identifying backdoor models as $c_r$ increases. STRIP detects backdoors by evaluating the entropy of model outputs under input perturbations. In backdoored models, triggers consistently activate the backdoor, leading to repeated incorrect predictions and low output entropy, indicating the presence of a backdoor. However, \methodname~camouflaging significantly reduces the ASR, meaning triggered inputs do not consistently produce misclassifications. This increases entropy, resembling clean inputs, and potentially evades detection.

\begin{figure}[!t]
    \centering
    \begin{subfigure}{0.48\linewidth}
        \centering
        \includegraphics[width=\linewidth]{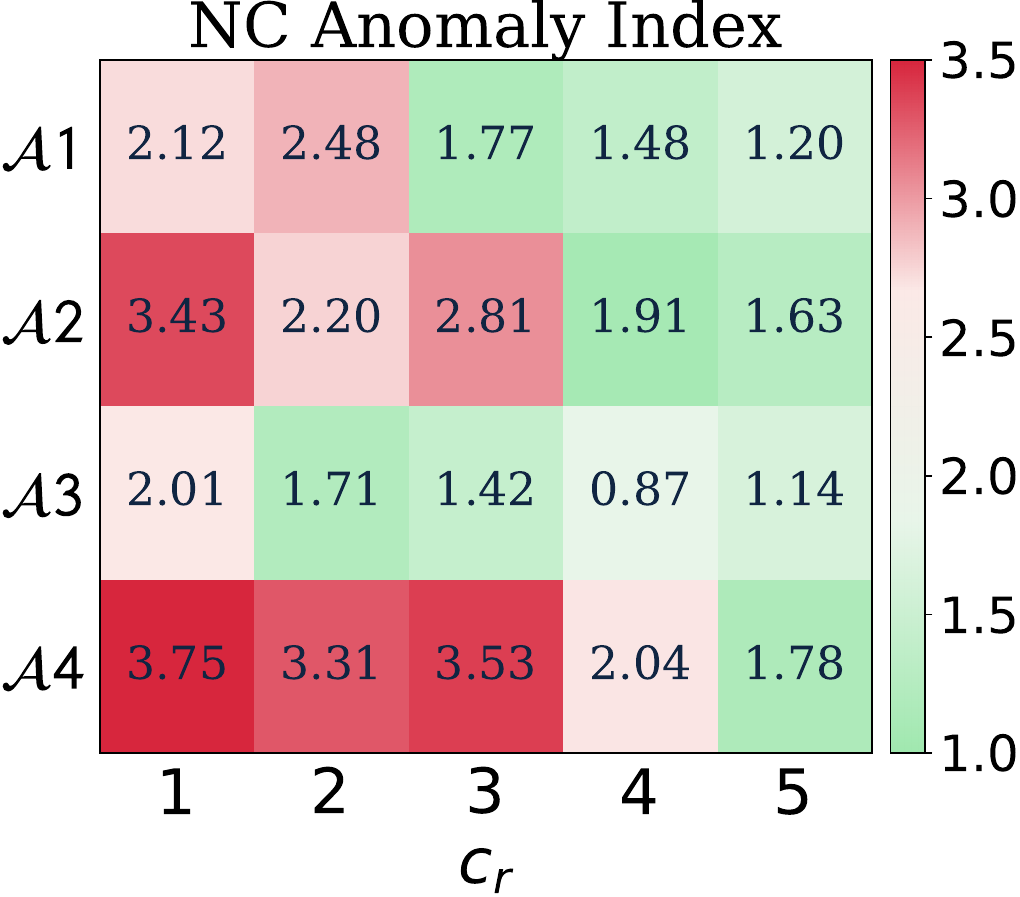}
        \caption{CIFAR10}
    \end{subfigure}\hspace{0.1cm}
    \begin{subfigure}{0.48\linewidth}
        \centering
        \includegraphics[width=\linewidth]{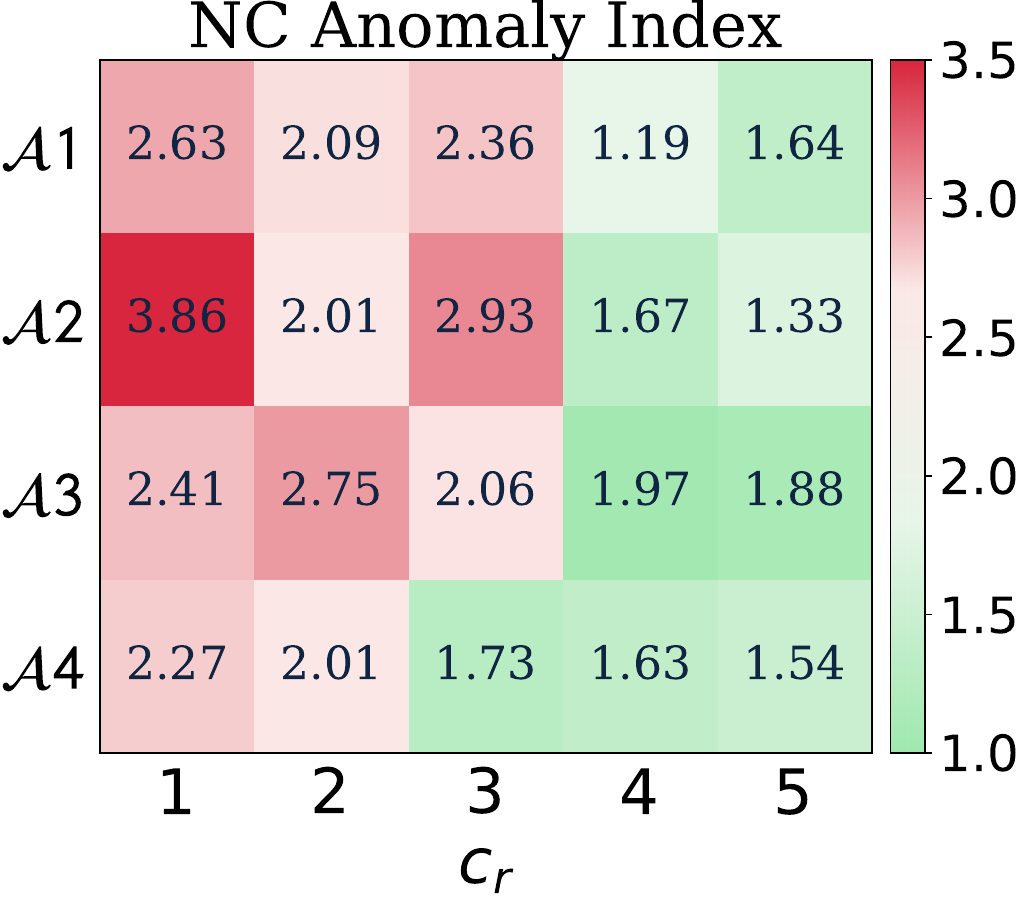}
        \caption{GTSRB}
    \end{subfigure}
    \begin{subfigure}{0.48\linewidth}
        \centering
        \includegraphics[width=\linewidth]{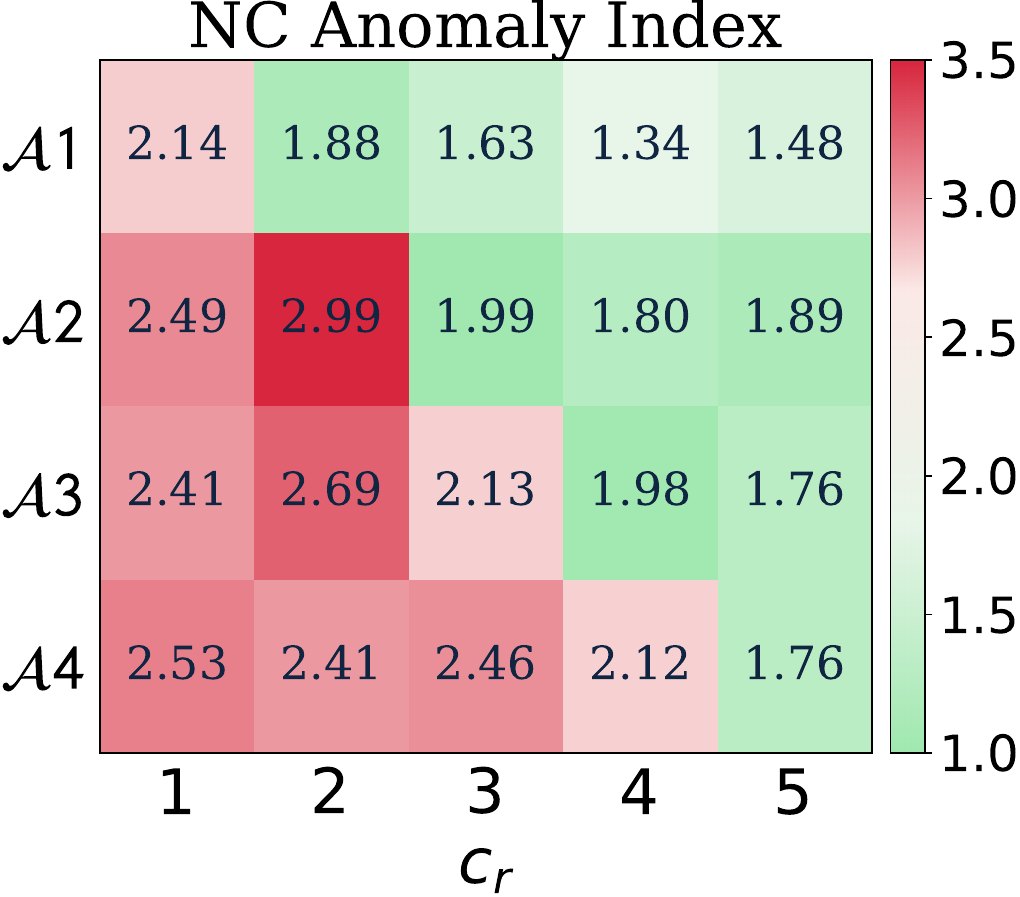}
        \caption{CIFAR100}
    \end{subfigure}\hspace{0.1cm}
    \begin{subfigure}{0.48\linewidth}
        \centering
        \includegraphics[width=\linewidth]{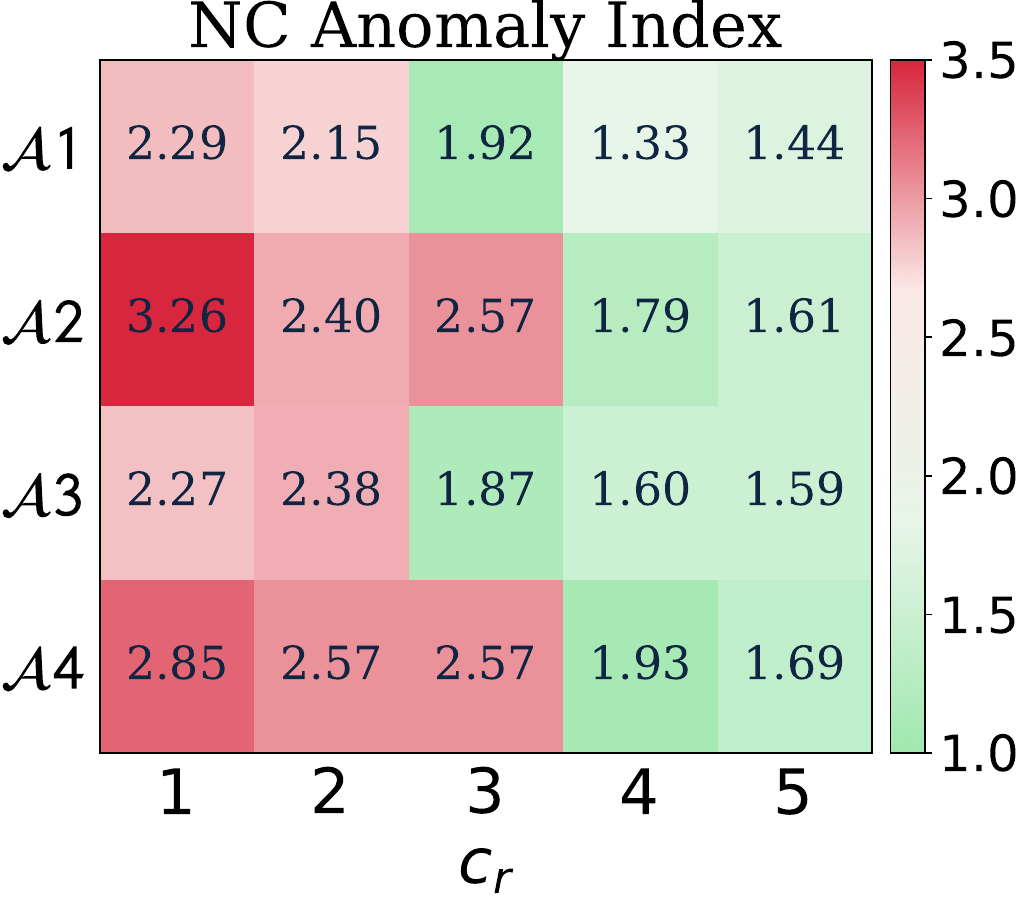}
        \caption{Tiny}
    \end{subfigure}
    \caption{Evaluation of \methodname~against NC for different datasets and attack methods. The anomaly index \textbf{greater than or equal to two} $(\geq 2)$ signifies the presence of backdoor in the model.}
    \label{fig:nc}
\end{figure}

\vspace{0.15cm}
\noindent \textbf{Neural Cleanse~\cite{nc} Defense Evaluation:} Figure~\ref{fig:nc} shows the performance of \methodname~camouflaging against the Neural Cleanse (NC) backdoor detection method for different attacks and datasets across varying $c_{r}$ under the setting of $\sigma = 10^{-3}$.
In NC evaluation, the \textit{NC Anomaly Index} is used to determine the presence of a backdoor, where a \textit{value greater than or equal to two indicates successful backdoor detection} and a value less than two signifies the backdoor is undetected. For CIFAR10 with $\mathcal{A}_1$, the NC anomaly index is 2.12 at $c_r = 1$, indicating the presence of a backdoor. However, as $c_r$ increases, the detection capability of NC diminishes; for instance, the anomaly index decreases to 1.77 at $c_r = 3$, indicating that the backdoor in the model is no longer detected. For GTSRB with $\mathcal{A}_1$, the anomaly index drops from 2.63 at $c_r = 1$ to 1.19 at $c_r = 4$. For CIFAR100 with $\mathcal{A}_1$, the anomaly index decreases from 2.14 at $c_r = 1$ to 1.88 at $c_r = 2$. Similarly, for Tiny with $\mathcal{A}_1$, the anomaly index decreases from 2.29 at $c_r = 1$ to 1.92 at $c_r = 3$. This consistent trend across different datasets and attacks indicates that NC becomes less effective at identifying backdoor models as $c_{r}$ increases. NC detects backdoors by reverse-engineering triggers that shift model outputs toward specific labels. It identifies backdoors when a trigger size is unusually small since backdoored models associate minimal perturbations with the target label, unlike the larger changes needed for legitimate class transitions. However, \methodname~camouflaging reduces the ASR, requiring larger triggers for misclassification. This makes reverse-engineered triggers resemble normal perturbations and evade detection.

\begin{figure}[!t]
    \centering
    \begin{subfigure}{0.48\linewidth}
        \centering
        \includegraphics[width=\linewidth]{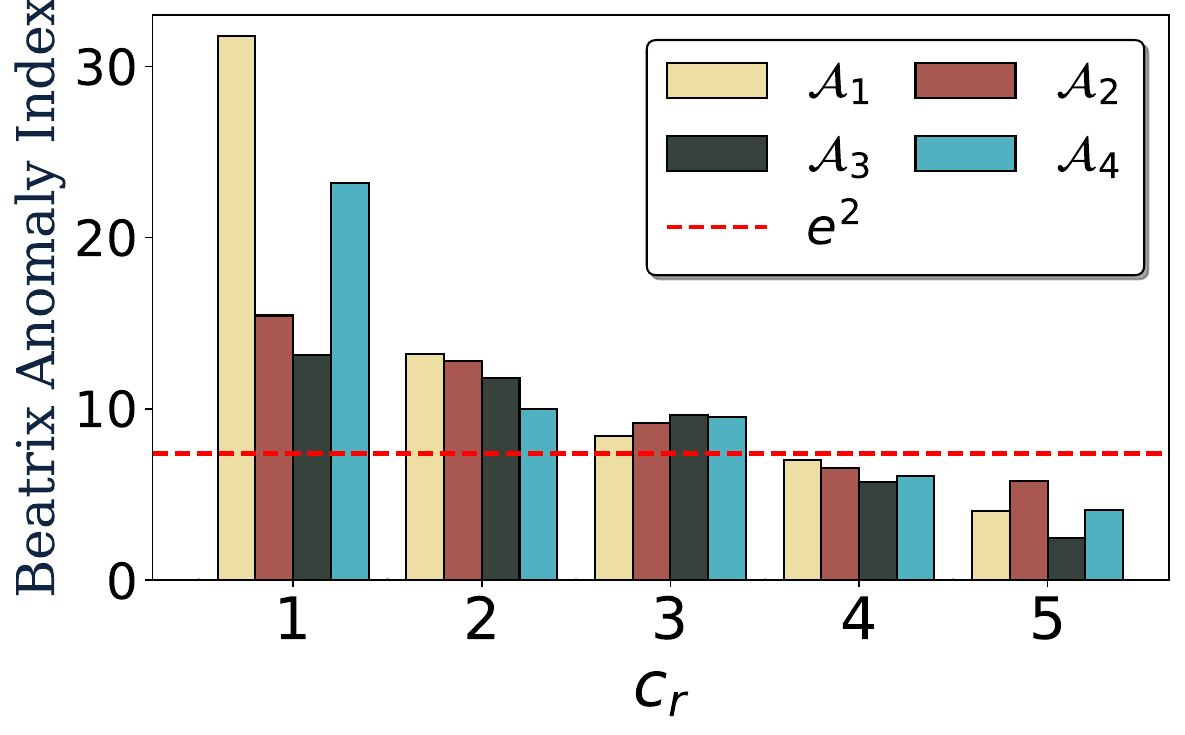}
        \caption{CIFAR10}
    \end{subfigure}\hspace{0.1cm}
    \begin{subfigure}{0.48\linewidth}
        \centering
        \includegraphics[width=\linewidth]{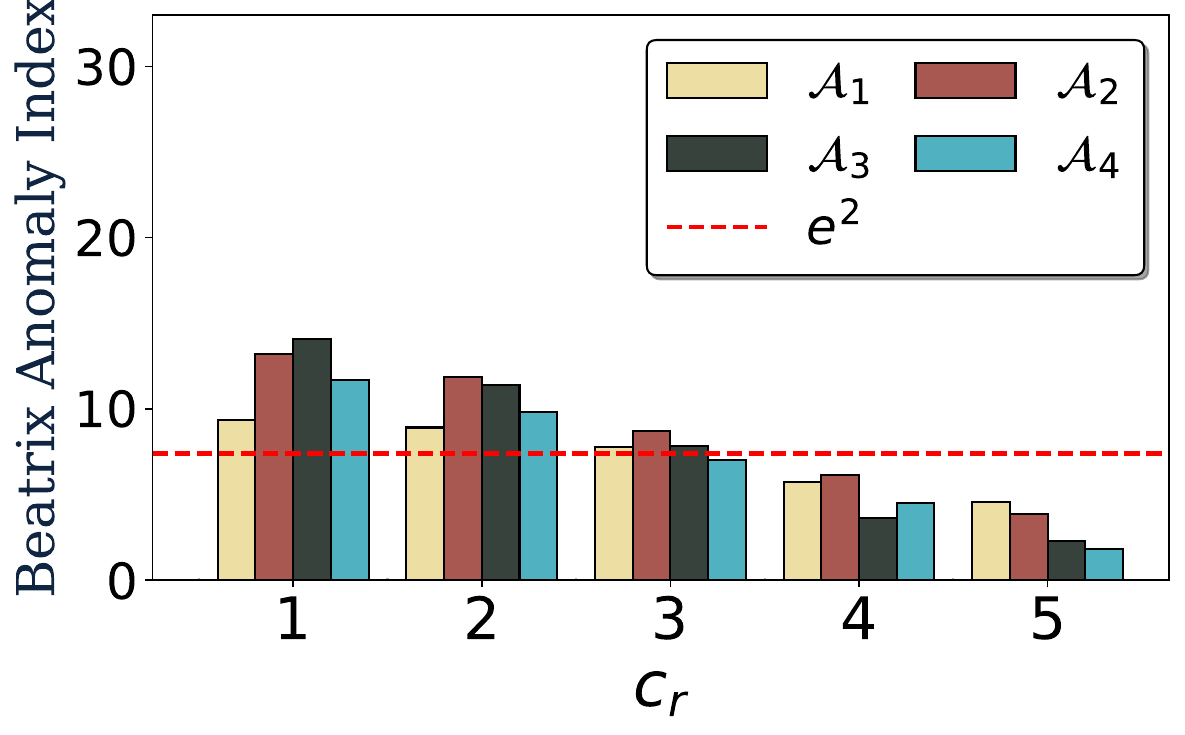}
        \caption{GTSRB}
    \end{subfigure}
    \begin{subfigure}{0.48\linewidth}
        \centering
        \includegraphics[width=\linewidth]{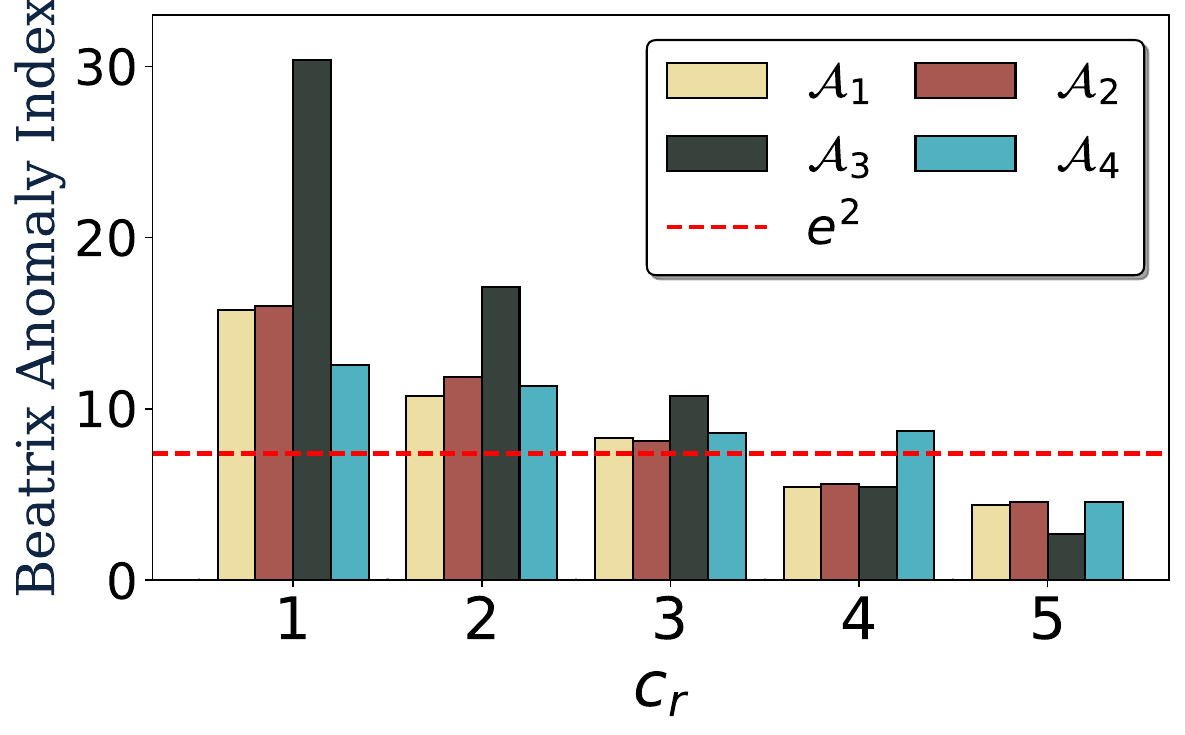}
        \caption{CIFAR100}
    \end{subfigure}\hspace{0.1cm}
    \begin{subfigure}{0.48\linewidth}
        \centering
        \includegraphics[width=\linewidth]{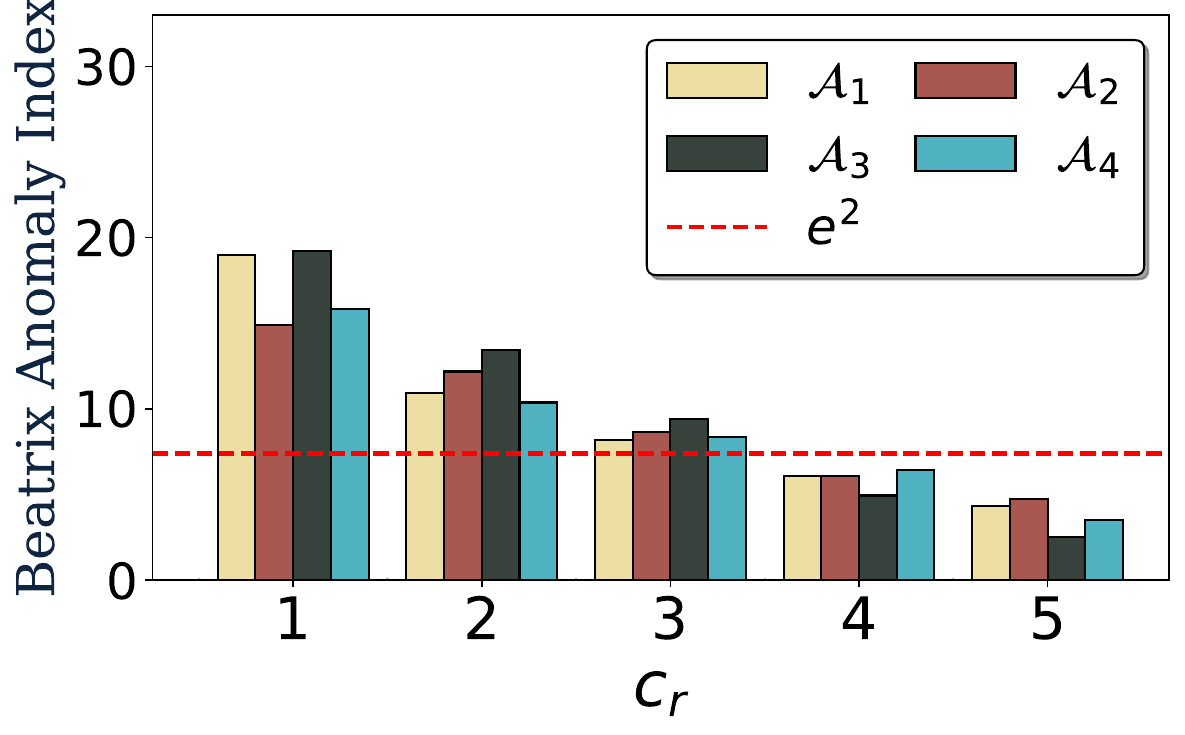}
        \caption{Tiny}
    \end{subfigure}
    \caption{Evaluation of \methodname~against Beatrix for different datasets and attack methods. The anomaly index \textbf{greater than or equal to} $e^2$ $(\geq 7.38)$ signifies the presence of backdoor in the model.}
    \label{fig:beatrix}
\end{figure}

\vspace{0.2cm}
\noindent \textbf{Beatrix~\cite{beatrix} Defense Evaluation:} Figure~\ref{fig:beatrix} shows the performance of \methodname~camouflaging against the Beatrix backdoor detection method for different attacks and datasets across varying $c_{r}$ under the setting of $\sigma = 10^{-3}$.
Similar to NC, in Beatrix evaluation, the \textit{Beatrix Anomaly Index} is used to determine the presence of a backdoor, where \textit{a value greater than or equal to $e^2\;(=7.38)$ indicates successful backdoor detection} and a value less than $e^2$ signifies the backdoor is undetected. For CIFAR10 with $\mathcal{A}_1$, the Beatrix anomaly index is 31.76 at $c_r = 1$, indicating the presence of a backdoor. However, as $c_r$ increases, the anomaly index decreases to 7.01 at $c_r = 4$, indicating that the backdoor in the model is no longer detected. For GTSRB with $\mathcal{A}_1$, the anomaly index drops from 9.37 at $c_r = 1$ to 5.75 at $c_r = 4$. For CIFAR100 with $\mathcal{A}_1$, the anomaly index decreases from 15.77 at $c_r = 1$ to 5.43 at $c_r = 4$. Similarly, for Tiny with $\mathcal{A}_1$, the anomaly index decreases from 18.97 at $c_r = 1$ to 6.06 at $c_r = 4$. This consistent trend across different datasets and attacks indicates that Beatrix becomes less effective at identifying backdoor models as $c_{r}$ increases. Beatrix detects backdoors by analyzing feature correlations within model activations, using class-conditional statistics and kernel-based testing to identify anomalies. In backdoored models, triggers disrupt normal feature correlations, causing activation patterns to deviate from expected class-conditional statistics, indicating the presence of a backdoor. However, \methodname~camouflaging significantly reduces ASR, meaning triggered inputs no longer consistently cause misclassifications. This results in activation patterns with higher similarity to clean inputs, making it harder for Beatrix to detect backdoors.

\section{Discussion and Future Work}
\noindent \textbf{Multi-Target Backdoor Attacks:}
Although our experiments focused on a single target attack, similar to other studies in the camouflage backdoor attack literature~\cite{DBLP:conf/nips/DiDA0S23,DBLP:conf/aaai/LiuWHM24,uba}, \methodname~can be readily adapted to more advanced multiple-target backdoor attacks~\cite{DBLP:journals/tdsc/XueHWL22}.

 \vspace{0.15cm}
\noindent \textbf{Approximate Unlearning:}
In our evaluation, we used the exact unlearning strategy~\cite{sisa}, but we believe \methodname~could also work with approximate unlearning methods~\cite{adaptive,amnesiac,unrollsgd,mcu,ermktp}. Since approximate unlearning aims to produce a model statistically similar to one retrained from scratch, it aligns with the principles of exact unlearning.

 \vspace{0.15cm}
\noindent \textbf{Potential Defense:}
The backdoor functionality is restored after unlearning requests are successfully executed. A naive defense against \methodname~could involve determining if unlearning requests are malicious by examining requested unlearning samples and the model's outputs.

\section{Conclusion}
This paper presents \methodname, a novel concealed backdoor attack targeting the data collection phase of the ML pipeline. Unlike existing methods, \methodname~requires no interaction with the target model or access to auxiliary data, enhancing its practicality. Experiments on four datasets and four trigger patterns show \methodname~significantly reduces ASR during pre-deployment and evades three popular backdoor detection methods. Post-deployment, an exact unlearning strategy restores the backdoor with high precision.

\vspace{0.15cm}
% \noindent{\textbf{Commitment to Open Science:}} We will open-source the implementation of \methodname.
\noindent \textbf{Acknowledgements:} This work has been supported by the NYUAD Center for Cyber Security under RRC Grant No. G1104.

\bibliographystyle{unsrt}
\bibliography{references}

\end{document}